\documentclass[aps,pre,onecolumn,floatfix,superscriptaddress,nofootinbib]{revtex4-1}
\usepackage{amsmath,amssymb,graphicx,cases}
\usepackage{epsfig}
\usepackage{textcomp}
\usepackage{epstopdf}
\usepackage{float}
\usepackage{braket}
\usepackage[normalem]{ulem}

\newcommand{\stkout}[1]{\ifmmode\text{\sout{\ensuremath{#1}}}\else\sout{#1}\fi}

\usepackage[colorlinks=true, urlcolor=blue, anchorcolor=blue, citecolor=blue,filecolor=blue,linkcolor=blue,menucolor=blue]{hyperref}

\begin{document}
\title{Area fluctuations on a subinterval of Brownian excursion}

\author{Baruch Meerson }
\email{meerson@mail.huji.ac.il}
\affiliation{Racah Institute of Physics, Hebrew University of
	Jerusalem, Jerusalem 91904, Israel}

\begin{abstract}
Area fluctuations of a Brownian excursion are described by the Airy distribution, which found applications in different
areas of physics, mathematics and computer science. Here we generalize this distribution to describe  the area fluctuations on a \emph{subinterval} of a Brownian excursion. In the first version of the problem (Model 1) no additional conditions are imposed. In the second version (Model 2) we study the distribution of the area fluctuations on a subinterval given the excursion area on the whole interval. Both versions admit convenient path-integral formulations. In Model 1 we obtain an explicit expression for the Laplace transform of the area distribution on the subinterval. In both models we focus on large deviations of the area by evaluating the tails of the area distributions, sometimes with account of pre-exponential factors.   When conditioning on very large areas in Model 2, we uncover two singularities in the rate function of the subinterval area fraction. They can be interpreted as dynamical phase transitions of second and third order.

\end{abstract}

\maketitle

\section{Introduction}

The Airy distribution describes fluctuations of the area under the curve, describing the position of a Brownian excursion as a function of time.
Since its first appearance nearly 40 years ago \cite{dar,Louch}, the Airy distribution has been observed in different areas of physics, mathematics, and computer science. One of the first applications of the Airy distribution was to inventory problems where it describes, for example, the distribution of the time spent by locomotives in a railway depot \cite{tac,tac2}. In graph theory this distribution describes fluctuations of the internal length of a rooted planar tree \cite{tac}. It also appears in the description of the computational cost of data storage algorithms \cite{comp1}.   In physics the Airy distribution appears  as the distribution of the maximum height of fluctuating interfaces in one dimension \cite{satyaprl,satcomt,solid}, the avalanche size distribution in some sandpile models \cite{sand}, the size fluctuations of ring polymers \cite{poly}, and even the position distribution of laser cooled atoms \cite{laser}. A brief review of some of these examples is presented in Ref.~\cite{satyacomputerreveiw}. Recently the Airy distribution was directly measured experimentally in a dilute colloidal system \cite{Agranov2020}.
Here we introduce and study two generalizations of the Airy distribution. Before we describe them, let us recap the main properties of the Airy distribution and its tails.

\begin{figure}[ht]
\includegraphics[width=0.40\textwidth,clip=]{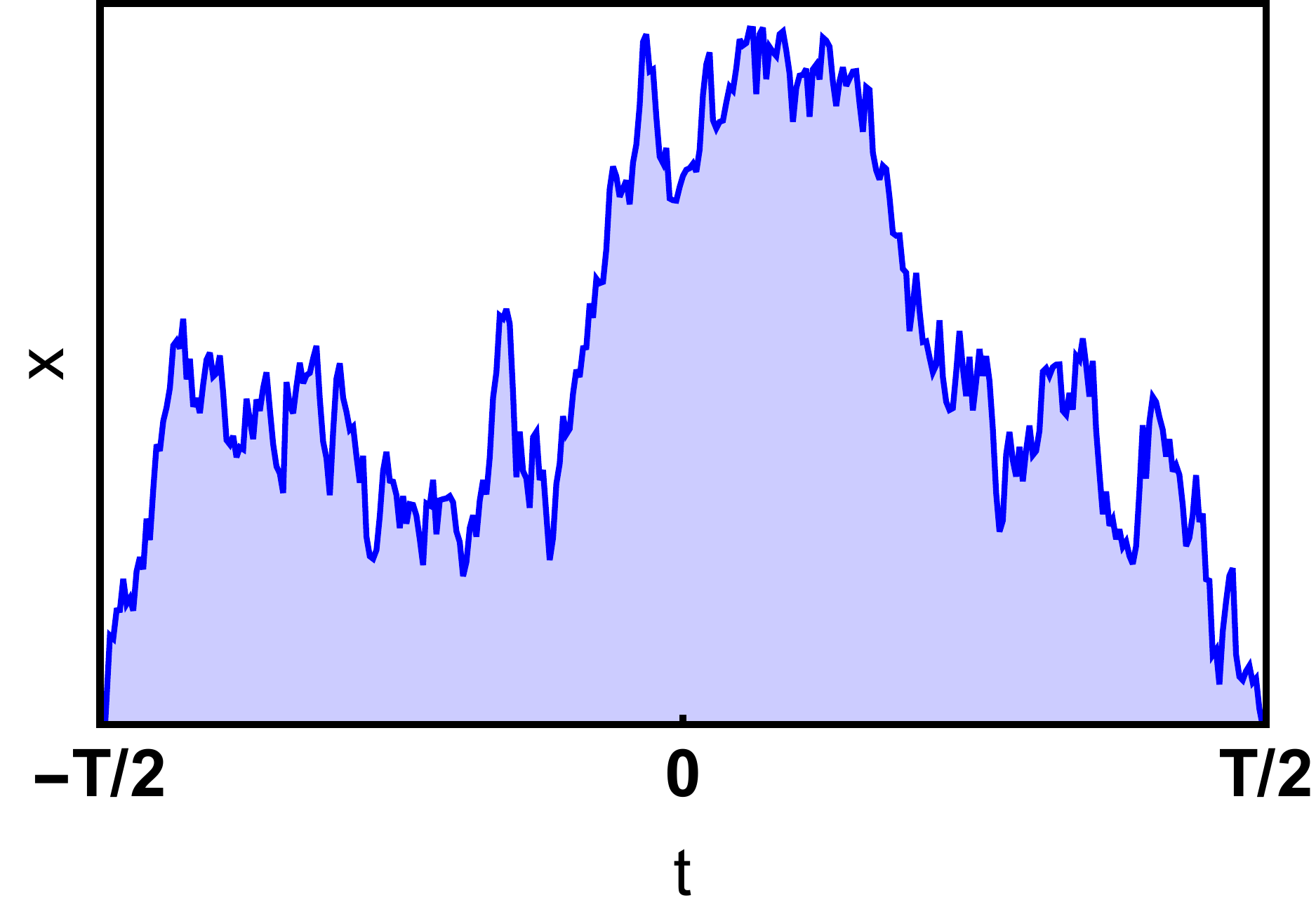}
\caption{A single realization of Brownian excursion on the time interval $|t|<T/2$. The Airy distribution describes fluctuations of the shaded area.}
\label{realizationAiry}		
\end{figure}

\begin{figure}[h]
\includegraphics[width=0.40\textwidth,clip=]{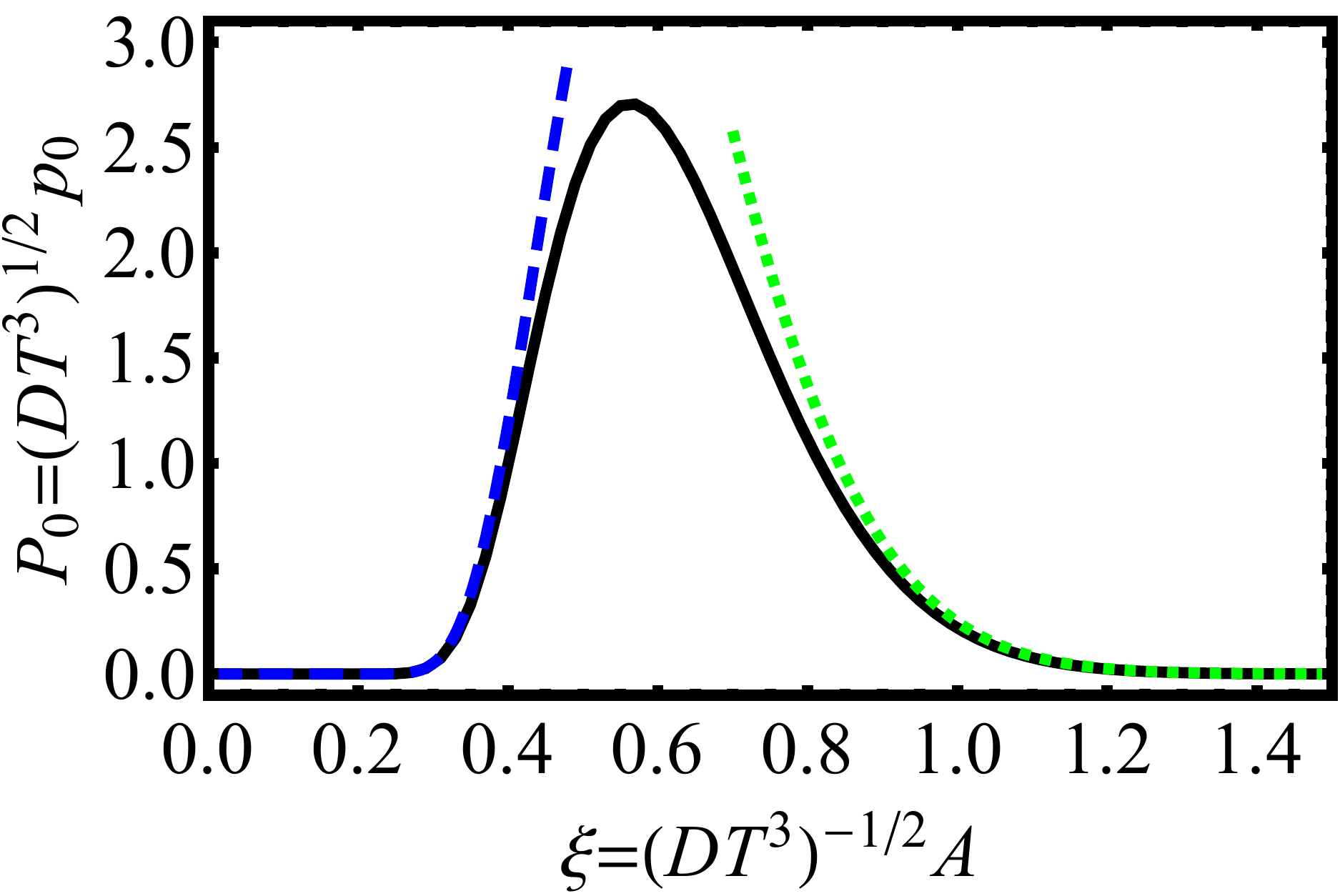}
\caption{The rescaled Airy distribution $p_0(A,T)$, see Eqs.~(\ref{AD1}) and Eq.~(\ref{AD2}). Shown is the scaling function $P_0(\xi)$ alongside with its asymptotics~(\ref{low}) and~(\ref{high}) for the left and right tails, respectively.}
\label{airyfig}		
\end{figure}

The Brownian motion $x(t)$, also known as the Wiener process,  is formally defined by the Langevin equation
$\dot{x}=\sqrt{D}\,\xi(t)$, where $\xi(t)$ is a delta-correlated zero-mean Gaussian noise, $\braket{\xi\left(t\right)\xi\left(t^{\prime}\right)}=\delta\left(t-t^{\prime}\right)$, and $D$ is the diffusion constant.  Here we consider a Brownian excursion: a Brownian motion conditioned to start and end very close to the origin, $x(t=-T/2)=x(t=T/2)=\epsilon>0$ (eventually, $\epsilon$ is sent to zero), and to stay positive\footnote{The condition that the process stays positive for all times $|t|<T/2$ distinguishes a Brownian excursion from a Brownian bridge. A bridge is allowed to cross the origin at $|t|<T/2$.}, $x(t)>0$ for all times $|t|<T/2$, see Fig.~\ref{realizationAiry}.
The area under the curve $x(t)$,
\begin{equation}
A=\int_{-T/2}^{T/2}x(t)dt ,\label{A}
\end{equation}
is a random variable, distributed according to the Airy distribution  \cite{dar,Louch,tac}
\begin{equation}
p_0(A,T)=\frac{1}{\sqrt{DT^3}}P_0\left(\frac{A}{\sqrt{DT^3}}\right), \label{AD1}
\end{equation}
where the scaling function
\begin{equation}
 P_0\left(\xi\right)=
 \frac{2\sqrt{6}}{\xi^{10/3}}\sum_{k=1}^{\infty}e^{-\beta_k/\xi^2}\beta_k^{2/3}
 U\left(-5/6,4/3,\beta_k/\xi^2\right),
 \label{AD2}
 \end{equation}
$U(\dots)$ is the confluent hypergeometric function \cite{hyper}, $\beta_k=2\alpha_k^3/27$, and $\alpha_k$ are the ordered absolute values of the zeros of the Airy function $\text{Ai}(\xi)$ \cite{airy}. The first values of $\alpha_k$ are $\alpha_1 = 2.33811\dots$, $\alpha_2=4.08795...$, $\alpha_3 = 5.52056\dots$, \textit{etc}.
The scaling function $P_0(\xi)$ is depicted in Fig. \ref{airyfig}. Also depicted there are the left and right tails of the distribution, which are described by the $\xi \ll 1$ and $\xi\gg 1$ asymptotics of $P_0(\xi)$ \cite{svante}:
\begin{numcases}
{{P_0(\xi)} \simeq}\frac{8\,\alpha_1^{9/2}}{81\, \xi^5}\,
 e^{-\frac{2\alpha_1^3}{27\xi^2}}, & $\xi\ll 1$, \label{low} \\
 \frac{72 \sqrt{6} \,\xi^2}{\sqrt{\pi}}\,e^{-6\xi^2},& $\xi\gg 1$. \label{high}
\end{numcases}
It has been shown recently \cite{Agranov2020} that the tails (\ref{low}) and (\ref{high}) can be described by two complementary methods of the theory of large deviations.  The small-area tail (\ref{low}) is described, up to the pre-exponential factor, by the Donsker-Varadhan formalism \cite{DonskerVaradhan,Ellis,hugo2009,Touchette2018}. In its turn, the large-area tail (\ref{high}) is captured by the optimal fluctuation method which, for a Brownian motion, becomes geometrical optics \cite{GF,majumdardasgupta,Ikeda2015,Holcman,Meerson2019,SmithMeerson2019a,SmithMeerson2019b,3short,Agranov2020,MMajumdar2020}.
For convenience of the reader, we will now briefly describe these findings of Ref. \cite{Agranov2020} .

The Donsker-Varadhan large deviation principle deals with
probability distributions of dynamical observables averaged over a long time.  In the context of the Airy distribution,
this dynamical observable is  the time-averaged position of the excursion, $\bar{X}=A/T$. It was observed in Ref.~\cite{Agranov2020} that, in the limit of $T \to \infty$ and $\bar{X}$ fixed,  the probability of observing a nonzero $\bar{X}$, as described by the Airy distribution, decays exponentially with the averaging time
\begin{equation}
-\ln{ P\left(\bar{X}\ll \sqrt{D T}\right)}\simeq T I\left(\bar{X}\right),\label{dv}
\end{equation}
thus obeying the Donsker-Varadhan large deviation principle \cite{DonskerVaradhan,Ellis,hugo2009,Touchette2018}. Then, from dimensional analysis, the rate function $I(\bar{X})$ must scale as $D/\bar{X}^2 = DT^2/A^2$. As a result,
$T I\left(\bar{X}\right) \sim DT^3/A^2 = 1/\xi^2$, reproducing (up to the pre-exponent) the scaling behavior \eqref{low} of the small -$A$ tail. To compute the numerical factor
$2\alpha_1^3/27$ in the exponent of Eq.~(\ref{low}), one should determine the ground state of a Schr\"{o}dinger-type ``tilted operator", obtained from the generator of the  Brownian excursion, constrained to have a specified area \cite{Agranov2020}. The trajectories $x\left(t\right)$, which dominate the small-$A$ tail (\ref{low}), stay in the vicinity of the origin, without crossing it,  for all times.  Importantly, the position distribution, corresponding to these trajectories, is stationary for most of the time \cite{Agranov2020}, which explains the simple exponential decay of $P$ with time ~\eqref{dv}.

The nature of the large-area tail (\ref{high}) is very different, as it is described by geometrical optics of diffusion \cite{Agranov2020}. This tail is dominated by a \emph{single} anomalously large excursion which realizes a specified area $A$ during a very short time $T$ (or, equivalently, a very large area during a specified time). This single most probable excursion $x_*(t)$  is called the
optimal path. It minimizes the Wiener's action
 \begin{equation}
 s\left[x\left(t\right)\right]=\frac{1}{2}\int_{-T/2}^{T/2} dt\,\dot{x}^2(t) ,\label{s1}
 \end{equation}
over all excursions $x\left(t\right)$ obeying the boundary conditions $x(t=-T/2)=x(t=T/2)=0$ and subject to the constraint (\ref{A}). The constraint can be accommodated via a Lagrange multiplier $\lambda$, defining an effective one-particle Lagrangian  $L\left(x, \dot{x}\right) = \dot{x}^2/2+\lambda x$.
The optimal path is a parabola,
 \begin{equation}
x_*\left(t\right)= \frac{3A}{2T}\left(1-\frac{4t^2}{T^2}\right) . \label{traj}
 \end{equation}
Then the large-$A$ tail (\ref{high}) is obtained, up to the pre-exponent, from the relation
$-\ln P\left[x_*\left(t\right)\right]\simeq -s[x_*(t)]/D$  \cite{Agranov2020}.

The two complimentary large deviation formalisms - the DV principle and the geometrical optics - are also at work in the tails of the two generalized distributions that we introduce and study in Sections~\ref{model1} and~\ref{model2}. Section~\ref{summary} includes a brief summary and discussion of our results.

\section{Model 1}
\label{model1}

\begin{figure}[ht]
			\includegraphics[width=0.40\textwidth,clip=]{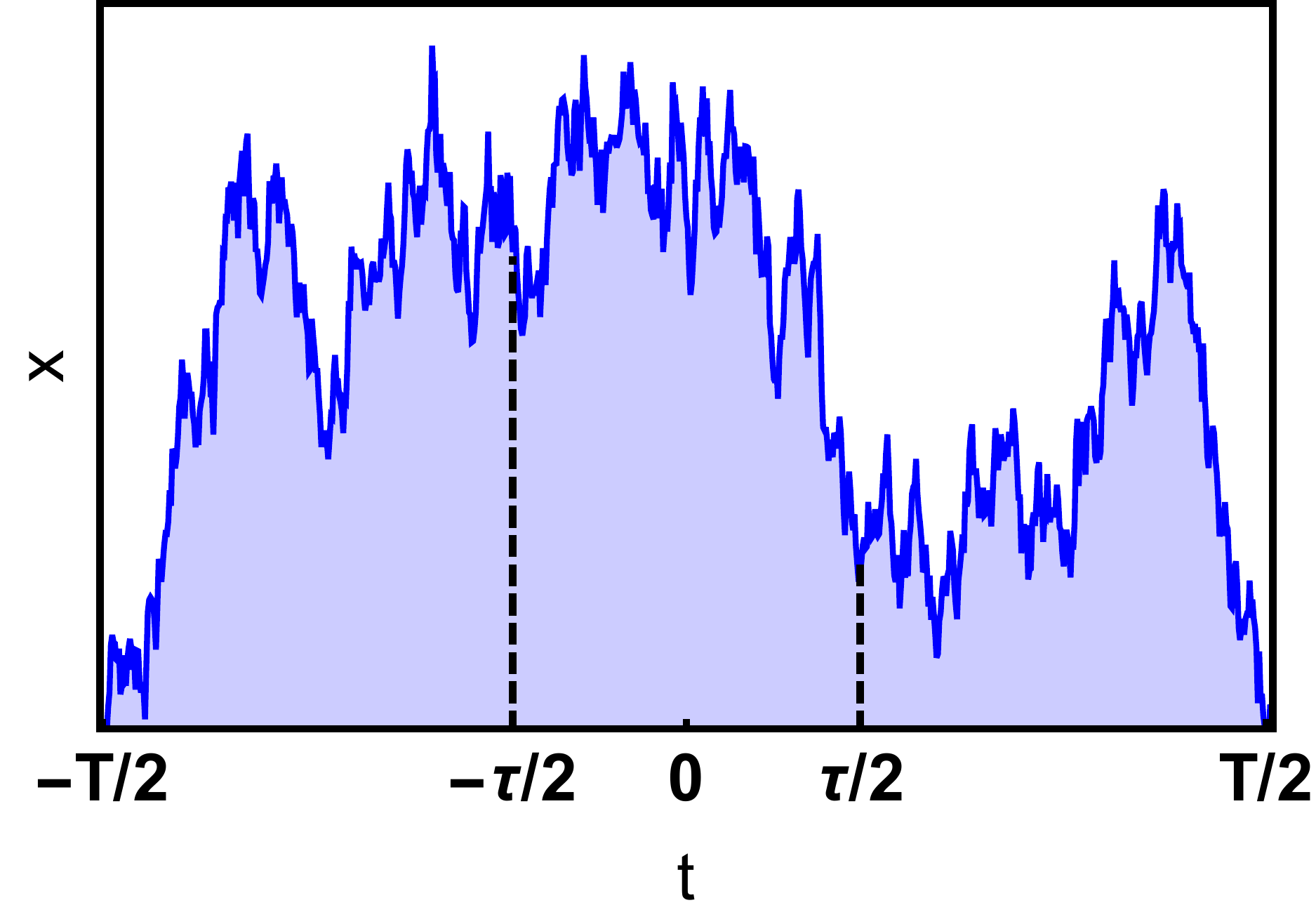}
\caption{A single realization of Brownian excursion on the time interval $|t|<T/2$. In Model 1 we study the probability distribution of fluctuations of the area $a$ on the subinterval $|t|<\tau/2$, without any additional conditions. In Model 2 we study the distribution of the area $a$ given the excursion area $A$ on the whole interval $|t|<T/2$.}
		\label{realization}		
\end{figure}

In some experiments one can collect data only from a subinterval of a Brownian excursion. To model this situation in a simple way,
let us consider a subinterval of time $|t|<\tau/2$, where $\tau\leq T$, and denote by $a$ the Brownian excursion area on this subinterval,
\begin{equation}
a=\int_{-\tau/2}^{\tau/2}x(t)\,dt , \label{a}
\end{equation}
see Fig. \ref{realization}. This area fluctuates from realization to realization, and we will study the probability distribution $p_a$ of these fluctuations \footnote{A similar extension of the Airy distribution was previously considered by Rambeau and Schehr \cite{Rambeau2009}. In their case the subinterval was adjacent to the starting point of the excursion. We will briefly discuss their model in Sec.~\ref{summary}.}. In Model 1 we do not impose any additional constraints. Dimensional analysis yields the following scaling behavior of $p_a$:
\begin{equation}\label{pa1}
p_a = \frac{1}{\sqrt{D T^3}}\, P\left(\frac{a}{\sqrt{D T^3}},\frac{\tau}{T}\right)\,,
\end{equation}
where $P(z_1,z_2)$ is a dimensionless function of the arguments $0<z_1<\infty$ and $0<z_2\leq 1$. In the particular case $\tau=T$ the distribution $p_a$ must coincide with the Airy distribution (\ref{AD1}) and (\ref{AD2}), so that $P(z_1,1)=P_0(z_1)$. For $\tau<T$, that is for $z_2<1$, $P(z_1,z_2)$ is not known. Here we will calculate its Laplace transform. We will also determine the tails of the area distribution $p_a$.

Let us rescale time $t/T\to t$ and the coordinate $x/\sqrt{DT} \to x$. As a result, $\tau/T \to \tau$ and   $a/\sqrt{DT^3} \to a$. In the rescaled units $p_a=P(a,\tau)$, so that $\int_0^{\infty} P(a,\tau) \,da = 1$.  The rescaled probability density $P(a,\tau)$ can be represented as
\begin{equation}\label{P(a)}
P(a,\tau) = \lim_{\epsilon\to 0} \frac{\int_0^{\infty}\int_0^{\infty}db_1 db_2 \Pi(\epsilon\to b_1) \Pi_a(b_1\to b_2)\Pi(b_2\to\epsilon)}{\Pi(\epsilon\to \epsilon)} .
\end{equation}
In Eq.~(\ref{P(a)}) we integrate over all possible intermediate points $b_1=x(-\tau/2)$ and $b_2=x(\tau/2)$ of the excursion $x(t)$ which start at $x=\epsilon>0$ at $t=-1/2$, arrive at $x=\epsilon$ at $t=1/2$ and obey the constraint (\ref{a}).
The probability density of a path passing through specified $b_1$ and $b_2$ is equal to the product of three probability densities (from now on we will call them simply probabilities): (i) the probability $\Pi(\epsilon\to b_1)$ to start at $x=\epsilon$ at $t=-1/2$ and arrive at $x=b_1$ at $t=-\tau/2$, (ii)  the probability $\Pi_a(b_1\to b_2)$ to start at $x=b_1$ at $t=-\tau/2$, arrive at $x=b_2$ and accumulate the area $a$, and (iii) the probability $\Pi(b_2\to \epsilon)$ to start at $x=b_2$ at $t=\tau/2$ and arrive at $x=\epsilon$ at $t=1/2$. This product of the three probabilities must be normalized by the probability $\Pi(\epsilon\to \epsilon)$ to start at $x=\epsilon$ at $t=-1/2$ and arrive at $x=\epsilon$ at $t=1/2$. As we are dealing with excursions, neither path is allowed to cross the origin.

Each of the probabilities  $\Pi(\epsilon\to b_1)$, $\Pi(b_2\to \epsilon)$ and $\Pi(\epsilon\to \epsilon)$ represents a particular case of the probability $\Pi(x_1\to x_2)$ of Brownian motion to start at $x=x_1>0$ at time $t=t_1$ and arrive at $x=x_2>0$ at time $t=t_2$ without crossing the origin for all $t_1<t<t_2$. This probability can be easily calculated by the image method \cite{Redner}:
\begin{equation}\label{pix1x2}
\Pi(x_1\to x_2)=\frac{e^{-\frac{(x_1-x_2)^2}{2
   (t_2-t_1)}}}{\sqrt{2\pi
   (t_2-t_1)}}-\frac{e^{-\frac{(x_1+x_2)^2
   }{2 (t_2-t_1)}}}{\sqrt{2\pi
   (t_2-t_1)}} .
\end{equation}

We now turn to the evaluation of the probability $\Pi_a(b_1\to b_2)$, which is conditioned on the area $a$. Following Ref. \cite{satcomt} (see also Refs. \cite{majumdardasgupta} and \cite{Rambeau2009}),  we represent $\Pi_a(b_1\to b_2)$ as a path integral:
\begin{equation}\label{piab1b2}
\Pi_a(b_1\to b_2) \,= \int\limits_{x(-\tau/2)=b_1}^{x(\tau/2)=b_2} {\mathcal D} x(t)\,\exp\left[-\frac{1}{2}\int\limits_{-\tau/2}^{\tau/2} \dot{x}^2(t) dt \right] \displaystyle\prod_{t=-\tau/2}^{\tau/2}\theta[x(t)] \,
\delta\left[\int\limits_{-\tau/2}^{\tau/2} x(t) dt-a\right] ,
\end{equation}
where the indicator function $\prod_{t=-\tau/2}^{\tau/2}\theta[x(t)]$ is equal to 1 if $x(t)>0$ for all $|t|<\tau/2$, and zero otherwise. The area constraint (\ref{a}) is taken into account by the delta-function. Let us consider the Laplace transform of $\Pi_a(b_1\to b_2)$:
\begin{equation}\label{Laplace}
\tilde{\Pi}_{\lambda} (b_1\to b_2) = \int_0^{\infty} \Pi_a(b_1\to b_2) e^{-\lambda a} da ,\quad \lambda \geq 0.
\end{equation}
Applying it to Eq.~(\ref{piab1b2}), we obtain
\begin{equation}\label{pilambda}
\tilde{\Pi}_{\lambda}(b_1\to b_2) \,= \int\limits_{x(-\tau/2)=b_1}^{x(\tau/2)=b_2} {\mathcal D} x(t)\,\exp\left[-\int\limits_{-\tau/2}^{\tau/2} \left(\frac{1}{2}\dot{x}^2(t)+\lambda x(t) \right) dt \right] \displaystyle\prod_{t=-\tau/2}^{\tau/2}\theta[x(t)] .
\end{equation}
This expression can be interpreted as the Euclidian propagator $\langle b_1 |e^{-\hat{H}\tau} |b_2\rangle$  for the Hamiltonian  $\hat{H} = -(1/2) d^2/dx^2+V(x)$ of a quantum particle in the potential
\begin{numcases}
{V(x)=}
\lambda x , & $x>0$, \nonumber \\
\infty ,& $x\leq 0$, \label{V(x)}
\end{numcases}
where the zero non-crossing condition is imposed by the infinite wall at $x=0$. This propagator was calculated in Ref. \cite{satcomt},
and we only present the results. At $\lambda>0$, the Hamiltonian $\hat{H}$ has only bound states, and the energy spectrum is discrete:
$E_k= 2^{-1/3} \alpha_k \lambda^{2/3}$, where $k=1,2,\dots$. To remind the reader,  $\alpha_k$'s are the absolute values of the zeros of the Airy function $\text{Ai}(z)$ on the negative real axis. The normalized eigenfunctions are given by \cite{satcomt}
\begin{equation}\label{wavefunctions}
\psi_k(x,\lambda) = \frac{(2\lambda)^{1/6}\text{Ai} \left[(2\lambda)^{1/3} x- \alpha_k \right]}{|\text{Ai}'(-\alpha_k)|},
\end{equation}
where $\text{Ai}'(z)=d \,\text{Ai}(z)/dz$, and we obtain
\begin{equation}\label{propagator}
\tilde{\Pi}_{\lambda}(b_1\to b_2) = \langle b_1 |e^{-\hat{H}\tau} |b_2\rangle = \sum\limits_{k=1}^{\infty} \psi_k(b_1,\lambda) \psi_k(b_2,\lambda) e^{-2^{-1/3}  \alpha_k \lambda^{2/3} \tau} .
\end{equation}

Now we apply the Laplace transform to Eq.~(\ref{P(a)}). $P(a,\tau)$ transforms into
$\tilde{P}(\lambda,\tau)$, and $\Pi_a(b_1\to b_2)$ transforms to $\tilde{\Pi}_{\lambda}(b_1\to b_2)$ given by Eq.~(\ref{propagator}). Using Eq.~(\ref{pix1x2}) and taking the limit of $\epsilon \to 0$, we obtain one of the central results of this work:
\begin{equation}\label{LaplaceP}
\tilde{P}(\lambda,\tau) = \frac{8\sqrt{2}}{\sqrt{\pi}\, (1-\tau)^3} \sum\limits_{k=1}^{\infty}
e^{-2^{-1/3}  \alpha_k \lambda^{2/3} \tau} [\mu_k(\lambda,\tau)]^2,
\end{equation}
where
\begin{equation}\label{Jk}
\mu_k(\lambda,\tau) = \int\limits_{0}^{\infty}  b \,e^{-\frac{b^2}{1-\tau}} \psi_k (b,\lambda)\,db,
\end{equation}
and $\psi_k$ are defined in Eq.~(\ref{wavefunctions}).

\begin{figure}[h]
\includegraphics[width=0.38\textwidth,clip=]{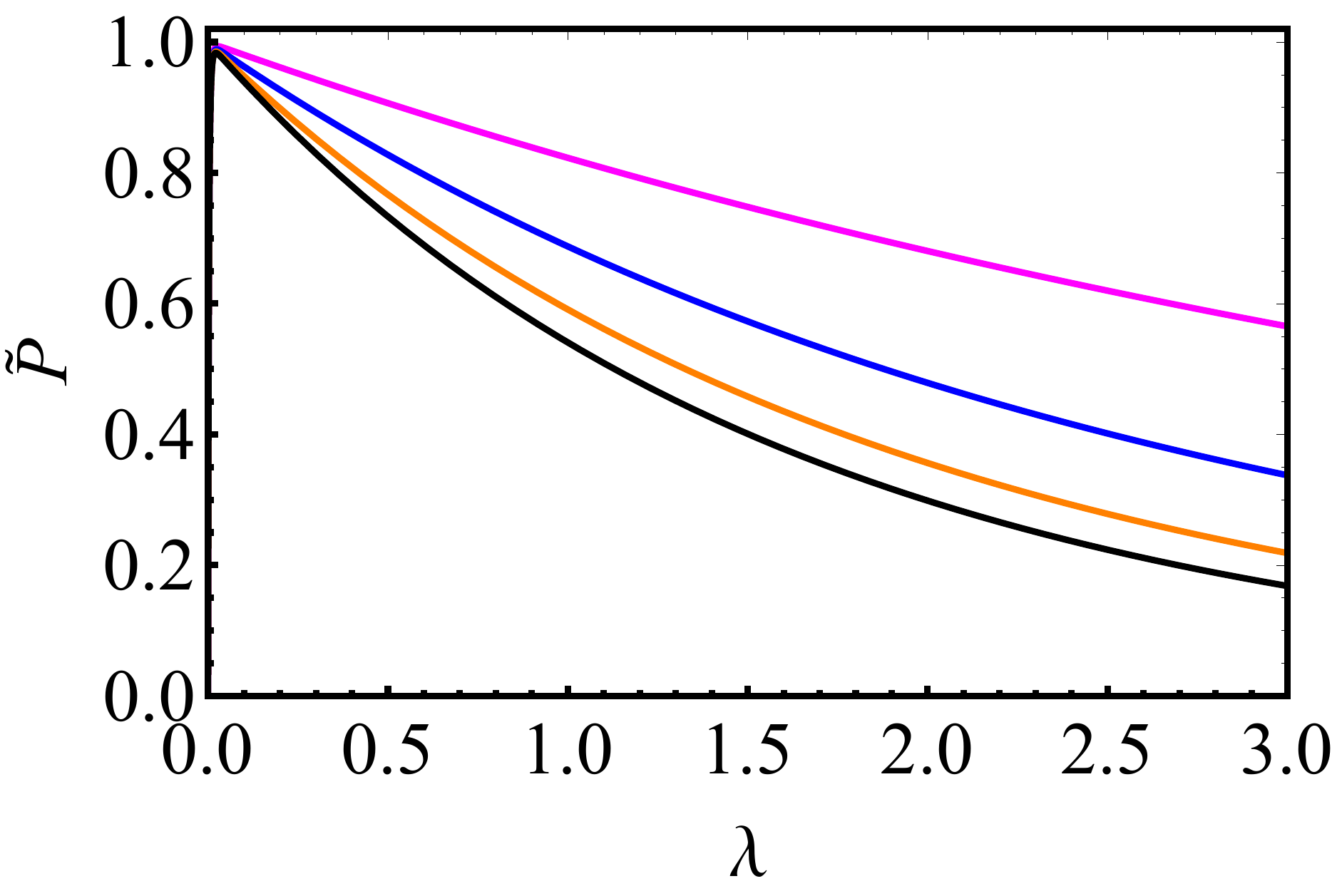}
\caption{The Laplace transform of the area distribution $P(a,\tau)$ on the subinterval $|t|<\tau/2$ for Model 1. Shown are numerical plots of $\tilde{P}(\lambda,\tau)$ as functions of $\lambda$ for $\tau=1/4$ (magenta), $1/2$ (blue) and $3/4$ (orange). Also shown, in black, is the Laplace transform~(\ref{LaplaceAD}) of the Airy distribution, $\tau=1$.}
\label{LTnumerical}		
\end{figure}

Equations~(\ref{LaplaceP}) and~(\ref{Jk}) describe exact Laplace transform of the probability distribution~$P(a,\tau)$. In order to restore the distribution itself, one should perform the inverse Laplace transform
\begin{equation}\label{inverseLaplace}
P(a,\tau)= \frac{1}{2\pi i} \int_{\gamma-i \infty}^{\gamma+i \infty} \tilde{P}(\lambda,\tau) e^{\lambda a} \,d\lambda.
\end{equation}
Figure \ref{LTnumerical} shows graphs of $\tilde{P}(\lambda,\tau)$ as a function of $\lambda$ for $\tau=1/4$, $1/2$ and $3/4$, obtained by numerical evaluation of Eqs.~(\ref{LaplaceP}) and~(\ref{Jk}). Also shown for comparison is the Laplace transform of the Airy distribution on the whole interval \cite{dar,Louch,tac,tac2,satcomt}:
\begin{equation}\label{LaplaceAD}
\tilde{P}_0(\lambda) = \sqrt{2\pi} \,\lambda \sum\limits_{k=1}^{\infty} e^{-2^{-1/3}\alpha_k\lambda^{2/3}} .
\end{equation}

As to be expected, Eq.~(\ref{LaplaceAD}) follows from our Eqs.~(\ref{LaplaceP}) and~(\ref{Jk}). Indeed,  in the limit of $\tau \to 1$, when the subinterval $|t|<\tau/2$ is close to becoming the whole interval $|t|<1/2$, the Gaussian factor $e^{-b^2/(1-\tau)}$ in Eq.~(\ref{Jk}) suppresses the integration over $b$ already at very small $b$. Therefore, we can expand $\psi_k(b,\lambda)$ in a Taylor series around $b=0$. We have
\begin{equation}\label{Taylor}
\text{Ai}\left[(2\lambda)^{1/3}b-\alpha_k\right] = \text{Ai}(-\alpha_k)+\text{Ai}'(-\alpha_k)(2\lambda)^{1/3} b + \dots .
\end{equation}
The zeroth-order term vanishes, and the first-order term yields
\begin{equation}\label{Jkcheck}
[\mu_k(\lambda,\tau\to 1)]^2 \simeq \frac{\pi \lambda}{8} (1-\tau)^3 .
\end{equation}
Plugging this expression into Eq.~(\ref{LaplaceP}), we see that the factors $(1-\tau)^3$ cancel out, and the resulting expression coincides with that of Eq.~(\ref{LaplaceAD}).

\subsection{Left tail of $P(a,\tau)$}
\label{lefta}

For the Airy distribution $P_0(A)$, the large-deviation regime $A\to 0$ corresponds to $\lambda\to +\infty$. As a result, the infinite sum over $k$ in the Laplace transform (\ref{LaplaceAD}) is dominated by the term $k=1$, while the rest of the terms are exponentially small compared to it \cite{Agranov2020}. The same property holds for the left (small-$a$) tail of the distribution $P(a,\tau)$. Therefore, we will assume (and check \textit{a posteriori}) that $\lambda^{2/3}\tau \gg 1$ and focus on the $k=1$ term. When $\tau$  is very close to $1$, there are two different asymptotic regimes of this tail. To identify them, let us consider the integrand of Eq.~(\ref{Jk}). There are two characteristic scales of integration over $b$. The Gaussian factor $e^{-b^2/(1-\tau)}$ defines the length scale $b_\tau \sim \sqrt{1-\tau}$, whereas the eigenfunction $\psi_1(b,\lambda)$ defines the length scale $b_{\lambda} \sim \lambda^{-1/3}$, see Eq.~(\ref{wavefunctions}).

When $b_\tau\ll b_{\lambda}$, we can expand $\psi_1(b,\lambda)$ in a Taylor series around $b=0$, as in Eqs.~(\ref{Taylor})-(\ref{LaplaceAD}). In the leading order we obtain
\begin{equation}\label{LaplaceADk1}
\tilde{P}(\lambda,\tau) \simeq \sqrt{2\pi} \,
\lambda e^{-2^{-1/3}\alpha_1 \lambda^{2/3} \tau} ,
\end{equation}
This result is very similar to Eq.~(\ref{LaplaceAD}) for  the $A\to 0$ asymptotic of the Airy distribution, where $a=A$. The only difference is the factor $\tau$ in the exponent of Eq.~(\ref{LaplaceADk1}).  We now apply the inverse Laplace transform (\ref{inverseLaplace}). Employing the large parameter $\lambda\gg1$, we can evaluate the integral over $\lambda$ in the complex plane by the method of steepest descent \cite{steepestdescent}. The saddle point $\lambda=\lambda_*$ is
the minimum point of the real function $\phi(\lambda) = \lambda a - 2^{-1/3} \alpha_1 \lambda^{2/3} \tau$ at $\lambda>0$, and we obtain $\lambda_*=4\alpha_1^3 \tau^3/27 a^3\gg 1$. The steepest-descent contour passes through the saddle point $(x=\lambda_*,y=0)$ of the complex plane $\lambda=x+iy$ in the $y$-direction. The final result, including the pre-exponential factor, is
\begin{equation}\label{intermediatea}
P(a,\tau) \simeq \frac{8 \alpha_1^{9/2}}{81 a^5}\, e^{-\frac{2 \alpha_1^3 \tau^3}{27 a^2}}\,,\quad \sqrt{1-\tau}\ll a\ll 1\,.
\end{equation}
As $\tau$ is very close to one, we can actually set $\tau=1$ in the exponent of Eq.~(\ref{intermediatea}), so this intermediate asymptotic does not depend on $\tau$ and, not surprisingly, coincides with the $A\to 0$ tail~(\ref{low}) of the Airy distribution, calculated by a different method \cite{svante}. The applicability domain of Eq.~(\ref{intermediatea}) can be understood as follows. The strong inequality $\lambda_*^{2/3}\tau \gg 1$ can be recast as $a\ll \tau^{3/2}$. On the other hand,  the strong inequality  $b_\tau\ll b_{\lambda}$ takes the form $a\gg \tau\sqrt{1-\tau}$.  The resulting double inequality
$\tau\sqrt{1-\tau} \ll a \ll \tau^{3/2}$ can be satisfied only if $\tau$ is very close to $1$. Due to this fact, the double inequality can be simplified to the form used in Eq.~(\ref{intermediatea}).

We now proceed to the more interesting opposite limit $b_\tau\gg b_{\lambda}$. Here the Gaussian factor  $e^{-b^2/(1-\tau)}$ in Eq.~(\ref{Jk}) can be replaced by $1$. The remaining integral can be evaluated, and we obtain
$\mu_1(\lambda,\tau) \simeq C\,\lambda^{-1/2},$ where
\begin{equation}\label{C}
C=\frac{\frac{\alpha_1}{27} \left(\frac{2\times 3^{2/3} \alpha_1^2 \,
   _1F_2\left(\frac{2}{3};\frac{4}{3},\frac{5}{3};-\frac{\alpha_1^3}{9}\right)}{\Gamma \left(\frac{7}{3}\right)}+\frac{3^{7/3} \alpha_1
   \, _1F_2\left(\frac{1}{3};\frac{2}{3},\frac{4}{3};-\frac{\alpha_1^3}{9}\right)}{\Gamma
   \left(\frac{2}{3}\right)}+9\right)-\text{Ai}'\left(-\alpha_1\right)}{\sqrt{2}\,
  \text{Ai}'\left(-\alpha_1\right)} = 2.29751\dots\,,
\end{equation}
and $ _1F_2(\dots)$ is the generalized hypergeometric function \cite{hyper}.
As a result,
\begin{equation}\label{Laplacek1}
\tilde{P}(\lambda,\tau)\simeq \frac{8 \sqrt{2} \,C^2}{\sqrt{\pi}\,\lambda  (1-\tau )^3}\,e^{-2^{-1/3} \alpha_1 \lambda ^{2/3} \tau}\,.
\end{equation}
As $\lambda^{2/3} \tau$ is a large parameter, the inverse Laplace transform (\ref{inverseLaplace}) can still be evaluated by the steepest-descent method, and the saddle point $\lambda_*$  is the same as before. Therefore, the leading exponential behavior is the same as in the asymptotic~(\ref{intermediatea}). The pre-exponential factor, however, is different:
\begin{equation}\label{smalla}
P(a,\tau) \simeq \frac{36 \sqrt{2}\,C^2\,a}{\sqrt{\pi}\alpha _1^{3/2} (1-\tau )^3 \tau ^{3/2}}\,e^{-\frac{2 \alpha _1^3 \tau ^3}{27 a^2}}, \quad a \ll \text{min}\left\{\tau^{3/2},\sqrt{1-\tau} \right\}\,.
\end{equation}
Here the presence of $\tau$ in the exponent is crucial, so this asymptotic strongly depends on $\tau$. The applicability domain of Eq.~(\ref{smalla}) is determined in the following way.
The condition $a\ll \tau^{3/2}$ is the same as in the asymptotic~(\ref{intermediatea}).
The condition $b_\tau\gg b_{\lambda}$ can be written as $a\ll \tau \sqrt{1-\tau}$.
Therefore, we must demand
\begin{equation}\label{doubletemporary}
a\ll \text{min} \left\{\tau^{3/2}, \tau\sqrt{1-\tau}\right\}\,.
\end{equation}
For $\tau$ very close to $1$ the condition $a\ll \tau\sqrt{1-\tau}$ can be simplified to $a\ll \sqrt{1-\tau}$. For $\tau\ll 1$ the condition~(\ref{doubletemporary}) becomes $a\ll \tau^{3/2}$. Finally, for $\tau$ not too close to $0$ or $1$, the condition is simply $a\ll 1$.
Altogether, we can simplify the double inequality~(\ref{doubletemporary}) to that of Eq.~(\ref{smalla}).

When $\tau$ is not close to $1$, the intermediate asymptotic regime~(\ref{intermediatea}) disappears, and the whole left tail is described by Eq.~(\ref{smalla}). Note that, to leading order (that is, neglecting the pre-exponential factors), the low-$a$ tail in both regimes of Eqs.~(\ref{intermediatea}) and (\ref{smalla}), is simply $-\ln P(a,\tau) \simeq (2 \alpha _1^3 \tau ^3)/(27 a^2)$ for $a\ll \tau^{3/2}$. This corresponds to the Donsker-Varadhan large deviation form
\begin{equation}\label{DVsmalla}
-\ln P(\bar{x}) \simeq \tau I(\bar{x}),
\end{equation}
where $\bar{x} = a/\tau$ is the time-averaged position of the excursion on the subinterval, and $I(\bar{x}) = 2\alpha_1^3/(27 \bar{x}^2)$. Notice that only the subinterval duration $\tau$ enters Eq.~(\ref{DVsmalla}), while the interval duration $T$ drops out.  That is,
up to pre-exponents, the low-$a$ tail is determined solely by the subinterval $|t|<\tau/2$. As we will see shortly, the large-$a$ tail is very different in this respect. The situation is also very different, in both tails, for Model 2.

\subsection{Right tail of $P(a,\tau)$}
\label{righta}
As in the case of the Airy distribution \cite{Agranov2020}, the $a\to \infty$ tail of $P(a,\tau)$ is dominated by a single, most likely excursion $x(t)$, which obeys the constraint of an unusually large area $a$. This large-deviation regime can be described by geometrical optics \cite{GF,majumdardasgupta,Ikeda2015,Holcman,Meerson2019,SmithMeerson2019a,SmithMeerson2019b,3short,MMajumdar2020}. In our problem the calculations involve
(i) a saddle-point evaluation of the path integral (\ref{pilambda}), and (ii) finding the optimal values of $b_1$ and $b_2$ so as to minimize the total action. This procedure is equivalent to determining the optimal path $x(t)$ on the whole interval $|t|<1/2$.
As one can check, the optimal path respects the time-reversal symmetry of the problem, so we can set $x(-t)=x(t)$ and $b_1=b_2\equiv b$.

The optimal trajectory on the subinterval $t<|\tau|/2$  can be found by minimizing the constrained action which appears inside the path integral~(\ref{pilambda}):
\begin{equation}\label{action1}
s[x(t)] = \int\limits_{-\tau/2}^{\tau/2} \left(\frac{1}{2}\dot{x}^2(t)+\lambda x(t) \right) dt\,,
\end{equation}
where $\lambda$ plays the role of a Lagrange multiplier. The resulting segment of the optimal trajectory is a parabola, symmetric with respect to $t=0$:
\begin{equation}\label{xinside1}
x\left(|t|<\frac{\tau}{2}\right) = b - \frac{\lambda}{2}\left(\frac{\tau^2}{4}-t^2\right),
\end{equation}
where the two constants $b=x(t=\pm \tau/2)$ and $\lambda$ are yet unknown.

\begin{figure}[h]
\includegraphics[width=0.40\textwidth,clip=]{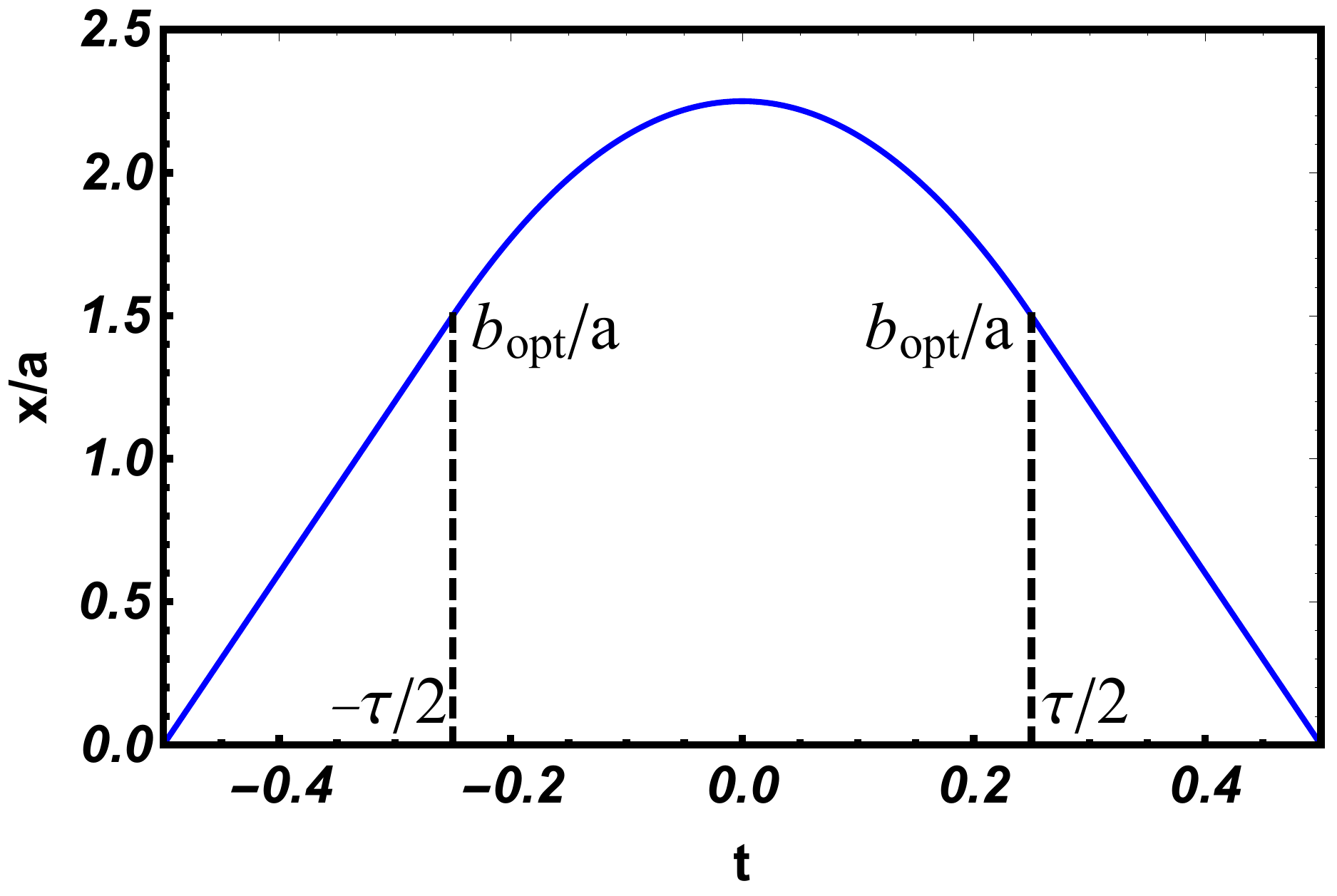}
\caption{Rescaled optimal path $x(t)/a$ of the Brownian excursion on the interval $|t|<1/2$, conditioned on the area $a$ on the subinterval $|t|<\tau/2$. In this example $\tau=1/2$.}
	\label{xonlya}		
\end{figure}

\begin{figure}[h]
\includegraphics[width=0.40\textwidth,clip=]{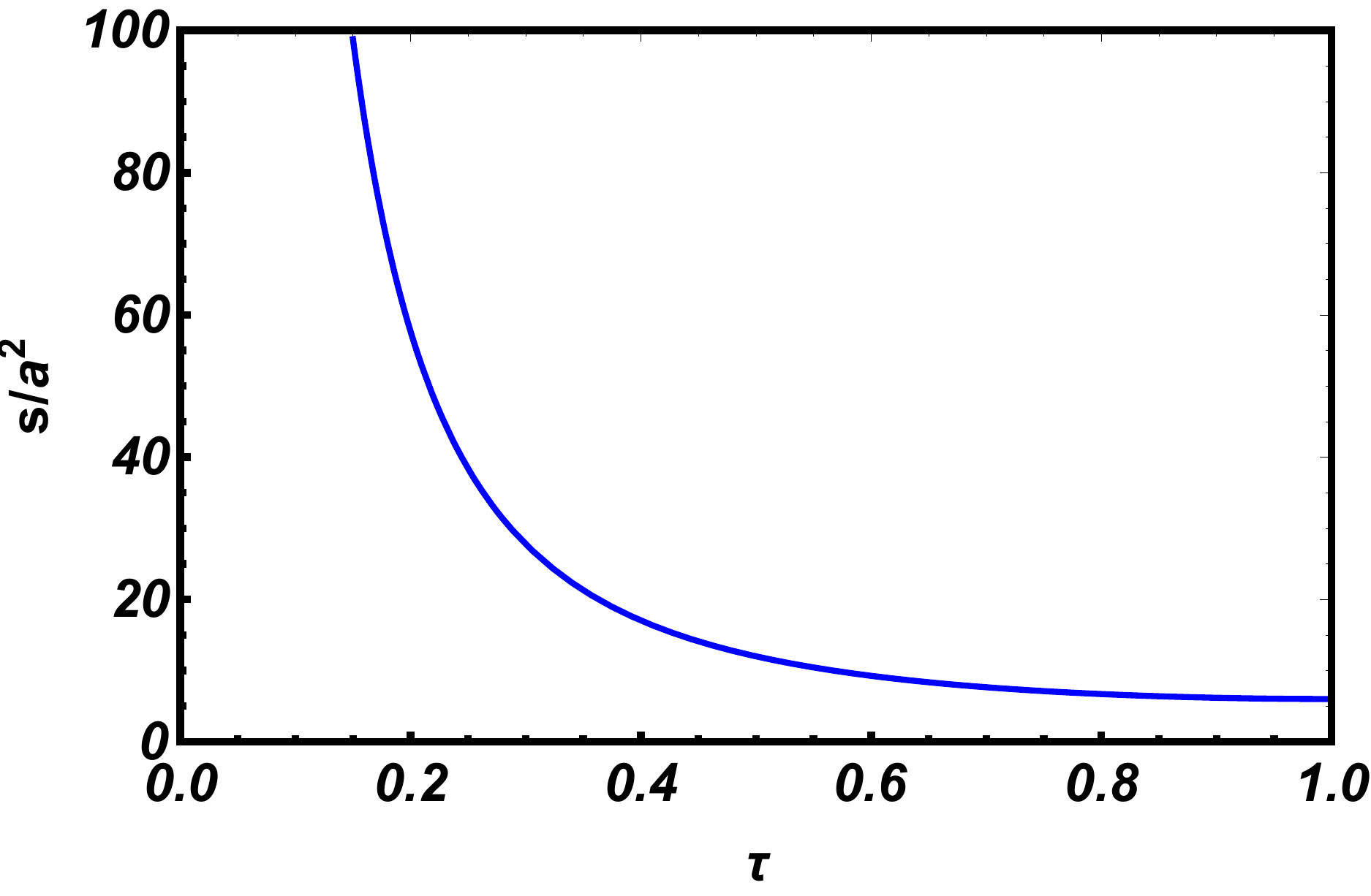}
\caption{Shown is the rescaled action $s(a,\tau)/a^2$ from Eq.~(\ref{righttailonlya}) as a function of the subinterval length $\tau$. As $\tau \to 0$, the function $s/a^2$ diverges as $2/\tau^2$.  At $\tau=1$ one obtains $s/a^2=6$, which corresponds to the Airy distribution, see Eq.~(\ref{high}).}
	\label{sovera2onlya}		
\end{figure}

The optimal path segments on the external subintervals $-1/2<t<\tau/2$ and $\tau/2<t<1/2$ minimize the \emph{unconstrained} action as described by Eq.~(\ref{action1}) with $\lambda=0$. Here the trajectories are ballistic:
\begin{equation}\label{xoutside1}
x(t) = b \frac{1-2|t|}{1-\tau}\,,
\end{equation}
where the conditions~$x(t=\pm 1/2)=0$ and the continuity conditions $x(\pm \tau/2)=b$ are already taken into account. The still unknown constants $b$ and $\lambda$ can be found from the constraint~(\ref{a}) and the continuity of $\dot{x}(t)$ at $t=\pm \tau/2$:
\begin{equation}\label{lambdab}
\lambda = -\frac{12 a}{\tau ^2 (3-2 \tau)},\quad b = b_{\text{opt}} \equiv \frac{3 a (1-\tau)}{\tau  (3-2 \tau)}
\end{equation}
An example of the resulting optimal trajectory is shown in Fig. \ref{xonlya}.
Now we can calculate the rescaled action:
\begin{equation}\label{action10}
s(a,\tau) = \frac{1}{2} \int_{-1/2}^{1/2} dt \,\dot{x}^2(t) .
\end{equation}
The contribution of the subinterval $|t|<\tau/2$ is
\begin{equation}\label{sinonlya}
s_{\text{in}} (a,\tau) = \frac{18 a^2}{\tau (3-2 \tau )^2}\,.
\end{equation}
The contribution of each of the two external subintervals, $-1/2<t<-\tau/2$ and $\tau/2<t<1/2$, is
\begin{equation}\label{soutonlya}
s_{\text{out}} = \frac{9 a^2 (1-\tau )}{\tau ^2(3-2 \tau )^2}\,.
\end{equation}
In contrast to the left tail of $P(a,\tau)$, the contributions of the internal and external subintervals are comparable.  Summing up all the three contributions, we obtain,  up to a pre-exponential factor, a Gaussian right tail of $P(a,\tau)$:
\begin{equation}\label{righttailonlya}
- \ln P(a,\tau) \simeq  s(a,\tau) =
\frac{6 a^2}{\tau ^2 (3-2 \tau )}\,.
\end{equation}
This asymptotic is accurate when the action $s(a,\tau)$ is much larger than unity. This condition reduces to the strong inequality $a\gg \tau$.  The function $s(a,\tau)/a^2$ versus $\tau$ is shown in Fig. \ref{sovera2onlya}.

When $\tau$ approaches zero, $s(a,\tau)$ diverges, and $P(a,\tau)$ rapidly vanishes, exhibiting an essential singularity $-\ln P(a,\tau) \sim 1/\tau^2$ as a function of $\tau$. This singularity is determined by the contribution $s_{\text{out}}$ of the outside subintervals.  When $\tau \to 1$, $s_{\text{out}}$ goes to zero, and the asymptotic~(\ref{righttailonlya}) reduces to that of the right tail of the Airy distribution, $s=6a^2$, as to be expected. It is useful to note that the contribution $s_{\text{out}}$ of each of the two external subintervals  to the action $s(a,\tau)$, Eq.~(\ref{soutonlya}), coincides in the leading order,  at $b\gg 1$,   with the logarithm of the probability $\Pi(\epsilon\to b)$, where $b$ is set to its optimal value from Eq.~(\ref{lambdab}). Indeed, setting $x_1=\epsilon$ and $x_2=b$ in the exact Eq.~(\ref{pix1x2})  and expanding the result at small $\epsilon$, we obtain
\begin{equation}\label{piepsb}
\Pi(\epsilon\to b) = \frac{4 \,\epsilon \, b}{\sqrt{\pi} (1-\tau)^{3/2}}\, e^{-\frac{b^2}{1-\tau}}.
\end{equation}
Plugging here the optimal value $b=b_{\text{opt}}$ from Eq.~(\ref{lambdab}), we see that the expression $b^2/(1-\tau)$ in the exponent coincides with Eq.~(\ref{soutonlya}).

Let us summarize our main results for Model 1.  The exact Laplace transform of the distribution $P(a,\tau)$ is described by Eqs.~(\ref{LaplaceP}) and~(\ref{Jk}). The small-$a$ tail of the distribution is described by two asymptotics: Eq.~(\ref{intermediatea}) for not too small $a$, and Eq.~\ref{smalla}) for the very small $a$. The leading-order result for both of these asymptotics coincides and is described by Eq.~(\ref{DVsmalla}). Finally, the large-$a$ tail of $P(a,\tau)$ is described, up to a pre-exponent, by Eq.~(\ref{righttailonlya}).
Now we proceed to Model 2, which turns out to be richer. In particular, it exhibits two dynamical phase transitions.

\section{Model 2}
\label{model2}

Even when the area distribution of the whole Brownian excursion is accessible in experiment, a sub-area distribution can still be of interest. Indeed, as we will see shortly, it provides an interesting additional characterization of the process. In Model 2 we again deal with a Brownian excursion on the interval $|t|<T/2$, see Fig. \ref{realization}. This time, however, we study the probability distribution  $p(a|A,\tau,T)$ of observing the subinterval area $a$, see Eq.~(\ref{a}), once the total area $A$  is specified by Eq.~(\ref{A}). The distribution  $p(a|A,\tau,T)$ is normalized to unity as follows: $\int_0^{\infty} p(a|A,\tau,T) da =1$. Clearly, $p=0$ for $a>A$.  The dimensional analysis yields the scaling form
\begin{equation}\label{pa2}
p(a|A,\tau,T) = \frac{1}{\sqrt{D T^3}}\, P\left(\frac{a}{\sqrt{D T^3}},\frac{A}{\sqrt{D T^3}},\frac{\tau}{T}\right)\,,
\end{equation}
where $P(z_1,z_2,z_3)$ is a dimensionless function of three dimensionless variables.  We will use the same dimensionless units as before, so that Eq.~(\ref{A}) becomes
\begin{equation}
A=\int_{-1/2}^{1/2}x(t)dt .\label{Arescaled}
\end{equation}

The rescaled probability density $P(a,A,\tau)$ can be written as
\begin{equation}\label{P(aA)}
P(a,A,\tau) = \lim_{\epsilon\to 0} \frac{\int_0^{\infty}\int_0^{\infty}db_1 db_2 \Pi'(\epsilon\to b_1) \Pi'_a(b_1\to b_2)\Pi'(b_2\to\epsilon)}{\Pi(\epsilon\to \epsilon) P_0(A)} .
\end{equation}
Here we integrate over all possible intermediate points $b_1=x(-\tau/2)$ and $b_2=x(\tau/2)$ of excursions $x(t)$ that start at $x=\epsilon>0$ at $t=-1/2$, arrive at $x=\epsilon$ at $t=1/2$ and obey the constraints (\ref{a}) and (\ref{Arescaled}). The probability of one such excursion is equal to the product of three probabilities: (i) the probability $\Pi'(\epsilon\to b_1)$ for the excursion to start at $x=\epsilon$ at $t=-1/2$ and arrive at $x=b_1>0$ at $t=-\tau/2$, (ii)  the probability $\Pi'_a(b_1\to b_2)$ to start at $x=b_1$ at $t=-\tau/2$, arrive at $x=b_2>0$ and accumulate the area $a$, and (iii) the probability $\Pi'(b_2\to \epsilon)$ to start at $x=b_2$ at $t=\tau/2$ and arrive at $x=\epsilon$ at $t=1/2$. The primes in the probabilities (i)-(iii) remind us of the
additional constraint on $A$ which modifies these probabilities compared with the probabilities $\Pi(\epsilon\to b_1)$, $\Pi_a(b_1\to b_2)$ and  $\Pi(b_2\to \epsilon)$ that we dealt with in Model 1.

The expression for $P(a,A,\tau)$ in Eq.~(\ref{P(aA)}) is normalized by the product of two probabilities: the probability $\Pi(\epsilon\to \epsilon)$ of the excursion to start at $x=\epsilon$ at $t=-1/2$, arrive at $x=\epsilon$ at $t=1/2$ and stay positive, and the probability $P_0(A)$ of the excursion to accumulate the area $A$. The former probability can be found from Eq.~(\ref{pix1x2}), the latter is nothing but the rescaled Airy distribution of the area $A$ on the interval $|t|<1/2$. We can represent the quantity
$$
q(a,A,\tau,b_1,b_2)\equiv\Pi'(\epsilon\to b_1) \Pi'_a(b_1\to b_2) \Pi'(b_2\to \epsilon)
$$
as a path integral:
\begin{equation}\label{piab1b2A}
q(a,A,\tau,b_1,b_2) = \int\limits_{x(-1/2)=\epsilon}^{x(1/2)=\epsilon} {\mathcal D} x(t)\,\exp\left[-\frac{1}{2}\int\limits_{-1/2}^{1/2} \dot{x}^2(t) dt \right] \displaystyle\prod_{t=-1/2}^{1/2}\theta[x(t)] \,
\delta\left[\int\limits_{-\tau/2}^{\tau/2} x(t) dt-a\right] \,\delta\left[\int\limits_{-1/2}^{1/2} x(t) dt-A\right],
\end{equation}
where the indicator function $\prod_{t=-1/2}^{1/2}\theta[x(t)]$ is equal to 1 if $x(t)>0$ for all $|t|<1/2$, and zero otherwise. The area constraints (\ref{a}) and (\ref{Arescaled}) are taken into account by two delta-functions. Applying to Eq.~(\ref{piab1b2A}) the double Laplace transform,
\begin{equation}\label{LaplaceA}
\tilde{q}(\lambda_1,\lambda_2,\tau,b_1,b_2) = \int_0^{\infty} \int_0^{\infty} q(a,A,\tau,b_1,b_2) e^{-\lambda_1 A - \lambda_2 a} dA\, da,\quad \lambda_{1,2} \geq 0 ,
\end{equation}
we obtain
\begin{equation}\label{pilambdaA}
\tilde{q}(\lambda_1,\lambda_2,\tau,b_1,b_2) = \int\limits_{x(-1/2)=\epsilon}^{x(1/2)=\epsilon} {\mathcal D} x(t)\,\exp\left[ - \int\limits_{-1/2}^{1/2} \left(\frac{1}{2}\dot{x}^2(t)+\lambda_1 x(t) \right) dt-\int\limits_{-\tau/2}^{\tau/2} \left(\frac{1}{2}\dot{x}^2(t)+\lambda_2 x(t) \right) dt\right] \displaystyle\prod_{t=-1/2}^{1/2}\theta[x(t)] .
\end{equation}
Regrouping the integrals in the square brackets, we can interpret $\tilde{q}(\dots)$ as the product of three Euclidian propagators,
\begin{equation}\label{threeprop}
\tilde{q}(\lambda_1,\lambda_2,\tau,b_1,b_2)=\langle \epsilon |e^{-\hat{H_1}(1/2-\tau/2)} |b_1\rangle\,
\langle b_1 |e^{-\hat{H_2}\tau}|b_2\rangle\,
\langle b_2 |e^{-\hat{H_1} (1/2-\tau/2)}|\epsilon\rangle\,,
\end{equation}
for the quantum Hamiltonians  $\hat{H}_{1,2} = -(1/2) d^2/dx^2+V_{1,2}(x)$.  The potential $V_1(x)$ acts on the external subintervals,  $\tau/2<|t|<1/2$, where it is equal to $V(x)$ from Eq.~(\ref{V(x)}) with $\lambda=\lambda_1$. The potential $V_2(x)$ acts on the subinterval $|t|<\tau/2$, where it is equal to $V(x)$ from Eq.~(\ref{V(x)}) with $\lambda=\lambda_1+\lambda_2$. For $\tau/2<|t|<1/2$ the eigenvalues are $E_k= 2^{-1/3} \alpha_k \lambda_1^{2/3}$, $k=1,2, \dots$, and the normalized eigenfunctions are described by Eq.~(\ref{wavefunctions}) with $\lambda=\lambda_1$.  For $|t|<\tau/2$ the eigenvalues are $E_k= 2^{-1/3} \alpha_k (\lambda_1+\lambda_2)^{2/3}$, and the eigenfunctions are described by Eq.~(\ref{wavefunctions}) with $\lambda=\lambda_1+\lambda_2$. Each of the three propagators in Eq.~(\ref{threeprop}) is an infinite series, similar to that of Eq.~(\ref{propagator}). Their product $\tilde{q}(\lambda_1,\lambda_2,\tau,b_1,b_2)$ is a triple series. This triple series should be integrated over $b_1$ and $b_2$ and inverse-double-Laplace transformed to yield the numerator of
$P(a,A,\tau)$ in Eq.~(\ref{P(aA)}). Although possible, these calculations are too cumbersome. Here we will calculate the leading-order asymptotic of $\tilde{q}$ at $\lambda_1\to \infty$ and $\lambda_2 \to \infty$. This calculation turns out to be quite simple, and it will ultimately enable us to determine, up to pre-exponential factors, the asymptotic behavior of the distribution $P(a,A,\tau)$ at small $a$ and $A$. After that we will calculate,
directly from Eqs.~(\ref{pa2}) and (\ref{piab1b2A}), the leading-order asymptotic of $P(a,A,\tau)$ at large $a$ and $A$.

\subsection{$P(a,A,\tau)$ at small $a$ and $A$}
\label{smallaA}

For $\lambda_1\gg 1$ and $\lambda_2\gg 1$ we can keep only the lowest term in the above-mentioned triple series. Up to pre-exponential factors, this gives
\begin{equation}\label{tildeq}
\tilde{q}(\lambda_1,\lambda_2,\tau) \sim \exp\left\{-2^{-1/3}\alpha_1 \left[ \lambda_1^{2/2}(1-\tau)+(\lambda_1+\lambda_2)^{2/3} \tau\right]\right\}.
\end{equation}
We suppressed the dependence of $\tilde{q}$ on $b_1$ and $b_2$ because, after the integration over $b_1$ and $b_2$, this dependence would affect only  pre-exponential factors that we are not trying to calculate. The inverse double Laplace transform reduces in this approximation to the double Legendre transform of the expression inside the exponent of Eq.~(\ref{tildeq}), where one should find the minimum point of a real function of two variables:
\begin{equation}\label{Phitwovar}
\Phi(\lambda_1,\lambda_2) = \lambda_1 A + \lambda_2 a -2^{-1/3}\alpha_1 \left[ \lambda_1^{2/2}(1-\tau)+(\lambda_1+\lambda_2)^{2/3} \tau\right]\,.
\end{equation}
The minimum is at the point
\begin{equation}\label{mintwovar}
\lambda_1=\frac{4 \alpha_1^3(1-\tau)^3}{27 (A-a)^3}, \quad \lambda_2 = \frac{4 \alpha_1^3 (A \tau-a
   ) \left(3 a^2 \tau ^2-3 a^2 \tau +a^2-3 a A \tau
   ^2+a A \tau +A^2 \tau ^2\right)}{27 a^3 (A-a)^3} ,
\end{equation}
and we can evaluate the numerator of Eq.~(\ref{P(aA)}): $q(a,A,\tau) \sim \exp [-\phi(\lambda_{1*},\lambda_{2*})]$. The function $\phi(\lambda_{1*},\lambda_{2*})$ can be presented in a symmetric form
\begin{equation}\label{phimintwovar}
\phi(\lambda_{1*},\lambda_{2*}) =
 \frac{2 \alpha_1^3}{27 A^2}\left[\frac{\tau^3}{f^2}+\frac{(1-\tau)^3}{(1-f)^2}\right] ,
\end{equation}
where we have introduced  the subinterval area fraction $f=a/A$, $0<f<1$.

In the denominator of Eq.~(\ref{P(aA)}) we can use the small-$A$ asymptotic~(\ref{low}) of the Airy distribution. We ignore both the pre-exponential factor in Eq.~(\ref{low}), and the factor $\Pi(\epsilon\to \epsilon)$ in its entirety, as they would only contribute to the pre-exponent of the distribution. Dividing
$\exp [-\phi(\lambda_{1*},\lambda_{2*})]$ by
$\exp (-2\alpha_1^3/27 A^2)$, we arrive at the following small-$a$ and small-$A$ asymptotic of the distribution $P(a,A,\tau)$:
\begin{equation}\label{ratefunctionr}
-\ln P(a,A,\tau) \simeq r(f,A,\tau) = \frac{2 \alpha_1^3}{27 A^2}\left[\frac{\tau^3}{f^2}+\frac{(1-\tau)^3}{(1-f)^2}-1\right] , \quad f = a/A .
\end{equation}
The large deviation function  $r(f,A,\tau)$ from Eq.~(\ref{ratefunctionr}) exhibits the characteristic $A^{-2}$ behavior of the left tail of the Airy distribution, as in Eq.~(\ref{low}). In addition, it has two important properties as a function of $f$ and $\tau$:
\begin{itemize}
\item {$r(f,A,\tau)$  vanishes at $f=\tau$, describing the simple fact that, at given $A$, the most probable value of $a$  is equal to $A$ times the relative length $\tau$ of the subinterval. Close to $f=\tau$ the rate function is quadratic in $f$, describing Gaussian fluctuations of $a$ around its the most probable value $\tau A$:
\begin{equation}\label{rgaussian}
 r(f,A,\tau)\simeq \frac{2 \alpha_1^3 (f -\tau)^2}{9  \tau  (1-\tau )}.
\end{equation}
}
\item{$r(f,A,\tau)$ obeys the symmetry relation $r(f,A,\tau) = r(1-f,A,1-\tau)$.}
\end{itemize}

To get an additional insight into these results, let us return to the original, dimensional variables and introduce three time-averaged particle positions:  $\bar{x}_{\tau}=a/\tau$  -- on the interval $(-\tau/2, \tau/2)$, $\bar{x}_{T-\tau}=(A-a)/(T-\tau)$ -- on the two disjoint intervals $(-T/2,-\tau/2)$ and $(\tau/2,T/2)$, and $\bar{X}=A/T$ -- on the whole interval  $(-T/2,T/2)$. Equation~(\ref{ratefunctionr}) takes the form
\begin{equation}\label{ratefunctionphysics}
-\ln P \simeq D\tau  \frac{2\alpha_1^3}{27 \bar{x}_{\tau}^2}+D(T-\tau) \frac{2\alpha_1^3}{27 \bar{x}_{T-\tau}^2}- DT  \frac{2\alpha_1^3}{27 \bar{X}^2}\,.
\end{equation}
Each of the three terms on the right hand side of this equation is proportional to the corresponding averaging time and describes three separate Donsker-Varadhan contributions [compare with Eq.~(\ref{dv})]. The last term has a minus sign, as we are dealing with a conditional probability.

\begin{figure}[h]
\includegraphics[width=0.38\textwidth,clip=]{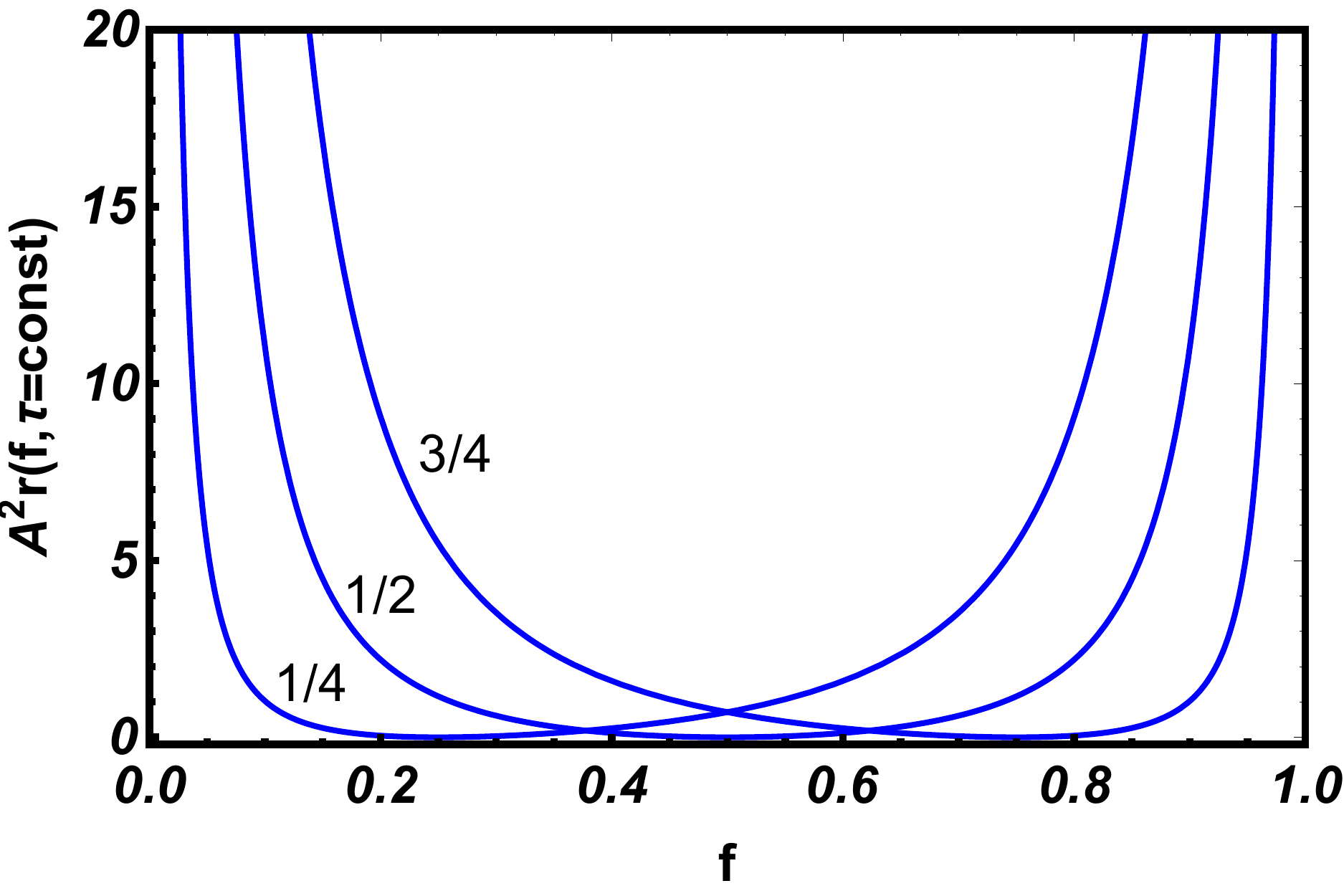}
\includegraphics[width=0.38\textwidth,clip=]{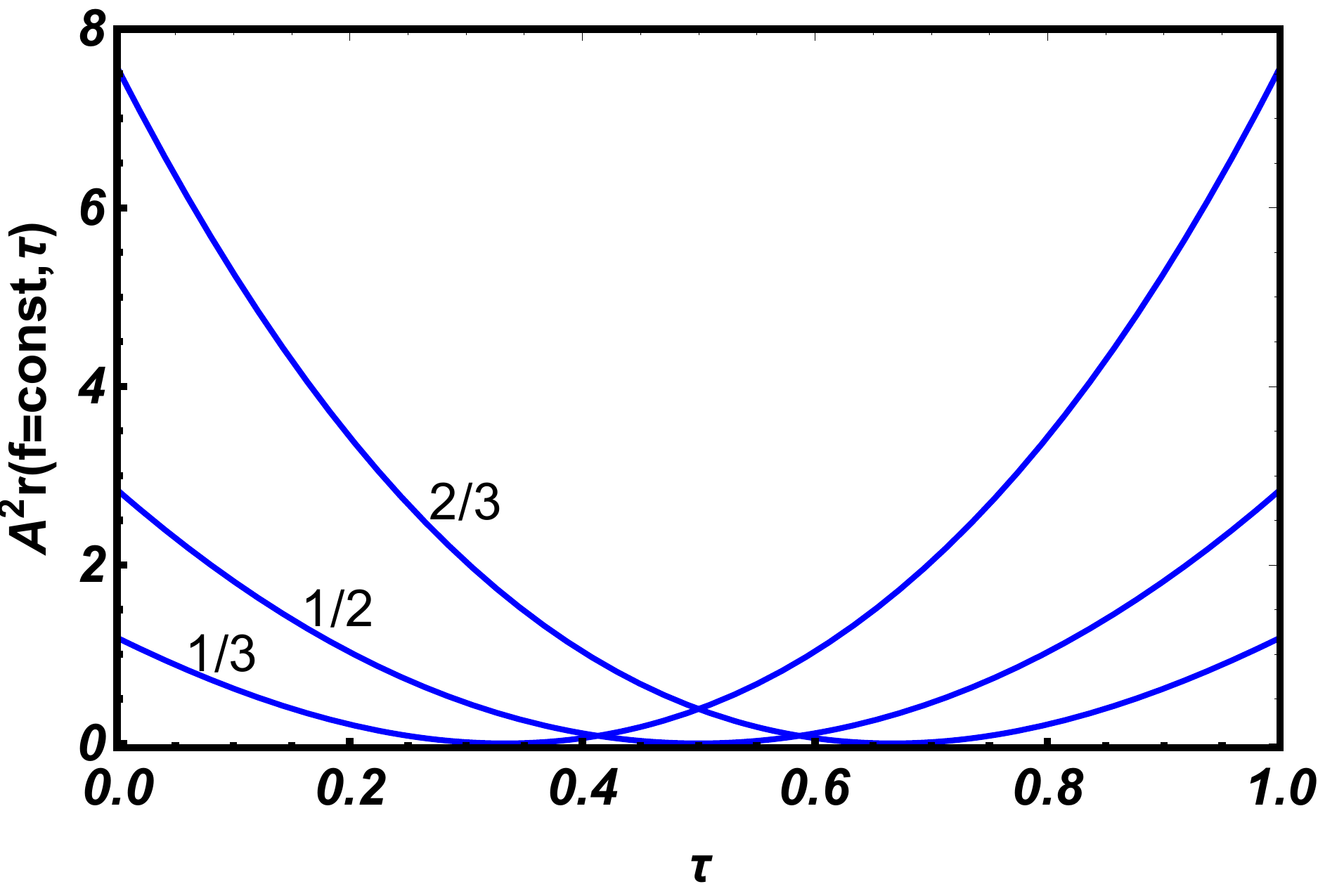}
\caption{The rate function $r(f,A,\tau)$  rescaled by $1/A^2$, as described by Eq.~(\ref{ratefunctionr}), is shown as a function of the area fraction $f=a/A$ at indicated fixed $\tau$ (the left panel) and as a function of $\tau$ at indicated fixed $f=a/A$ (the right panel).}
\label{rvsfandtau}		
\end{figure}

\subsection{$P(a,A,\tau)$ at large $a$ and $A$}
\label{largeaA}

\begin{figure}[h]
	\begin{tabular}{ll}
		\includegraphics[width=0.32\textwidth,clip=]{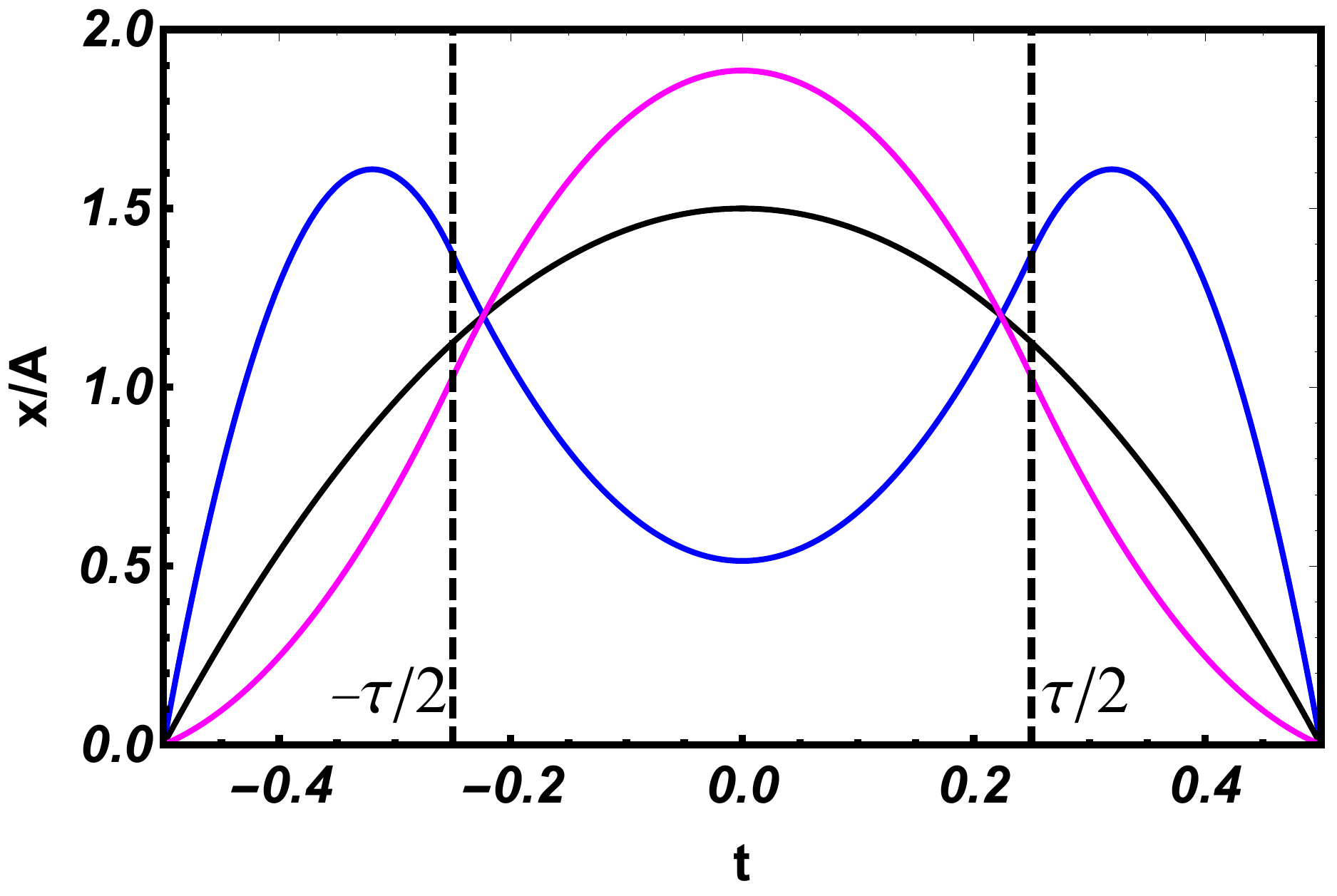}
		\includegraphics[width=0.32\textwidth,clip=]{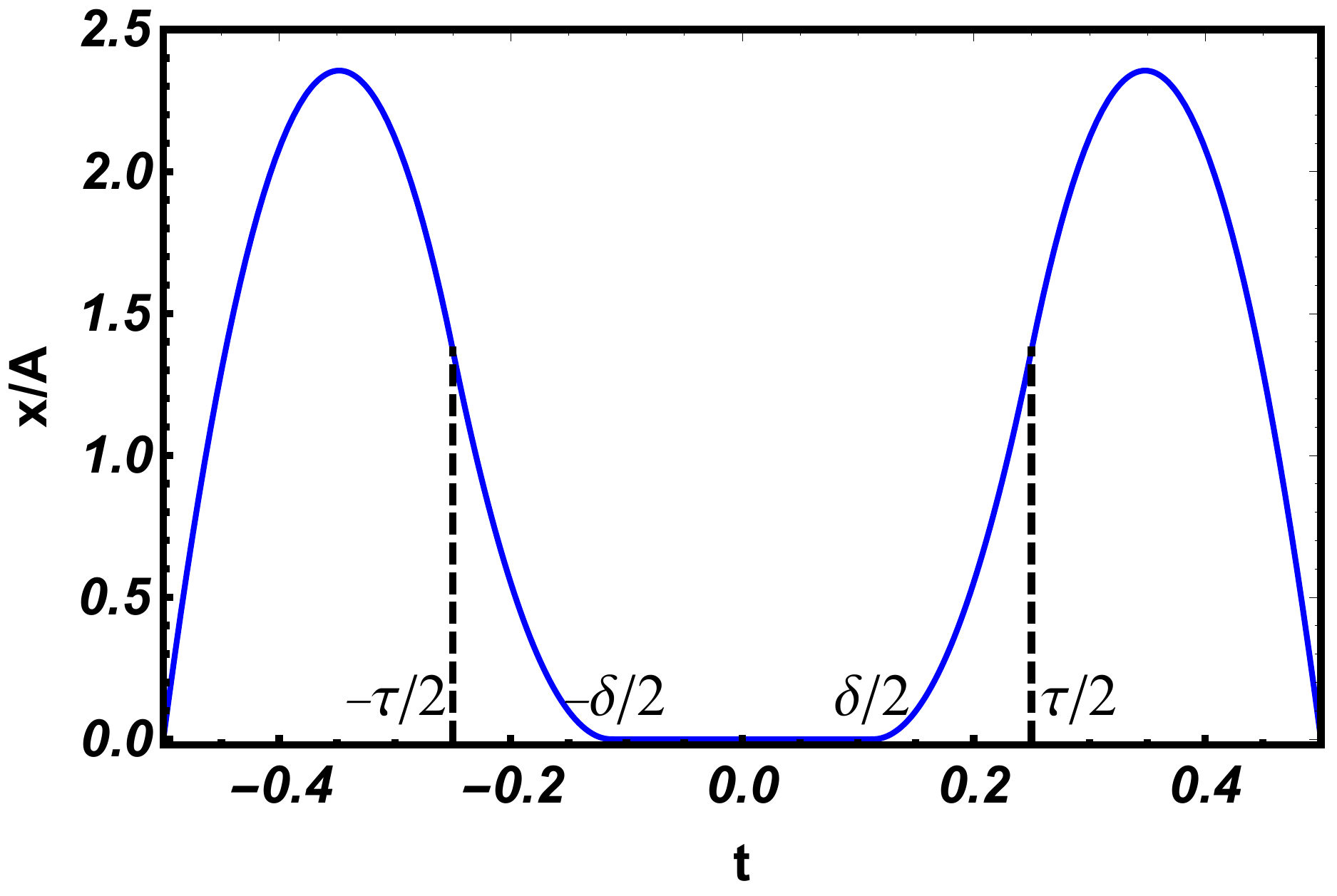}
        \includegraphics[width=0.32\textwidth,clip=]{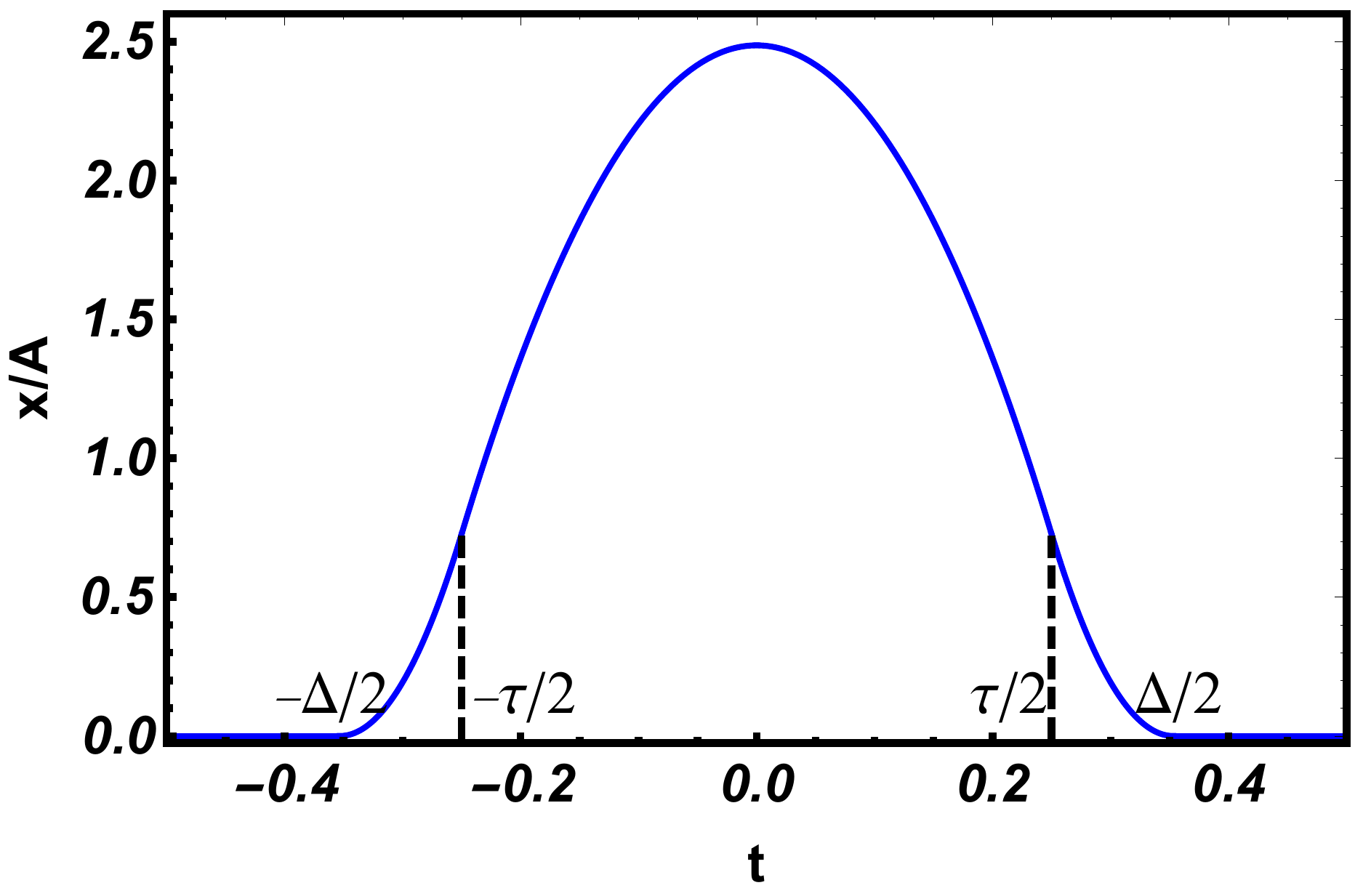}
	\end{tabular}	
	\caption{Optimal paths of the Brownian excursion  conditioned on large $A$ and $a=fA$ for $\tau=1/2$, so that $f_1=1/4$, $f_2=5/6$, $\delta=(3 - \sqrt{2})/7=0.2265\dots$, and $\Delta=0.7064\dots$. Left panel: three examples of the optimal paths as described by Eqs.~(\ref{xoutside}) and (\ref{xinside}) for $f=2/5$ (blue), $11/16$ (black) and $4/5$ (magenta), respectively. Middle panel: the optimal path as described by Eqs.~(\ref{xoutsideonevoid})-(\ref{delta}) for $f=1/8$.  Right panel: the optimal path  as described by Eqs.~(\ref{zerotwovoids})-(\ref{Delta}) for $f=0.95$.}
	\label{combined}		
\end{figure}

The limit of large $a$ and $A$ is described by geometrical optics. The optimal path $x(t)$ on the whole interval $|t|<1/2$ is constrained by the two conditions (\ref{a}) and (\ref{Arescaled}), which is accounted for by two Lagrange multipliers $\lambda_1$ and $\lambda_2$. As in Model 1, the solution includes the determination of the \textit{a priori} unknown optimal value of $b_1=b_2=b$. The constrained action to be minimized appears inside the path integral~(\ref{pilambdaA}):
\begin{equation}\label{sA}
s[x(t)] =  \int\limits_{-1/2}^{1/2} \left(\frac{1}{2}\dot{x}^2(t)+\lambda_1 x(t) \right) dt+\int\limits_{-\tau/2}^{\tau/2} \left(\frac{1}{2}\dot{x}^2(t)+\lambda_2 x(t) \right) dt .
\end{equation}
We decompose the interval $|t|<1/2$ into three subintervals, $-1/2<t<-\tau/2$, $|t|<\tau/2$ and $\tau/2<t<t/2$. The optimal path $x(t)$ consists of three parabolic segments. The parabolic segments on the external subintervals $\tau/2<|t|<1/2$ obey the boundary conditions $x(-1/2)=x(1/2)=0$. The adjacent parabolas match each other together with their first derivatives at $|t|=\tau/2$. Finally, the solution obeys the area constraints  (\ref{a}) and (\ref{Arescaled}). Altogether these conditions yield all the integration constants and the Lagrange multipliers $\lambda_1$ and $\lambda_2$ uniquely and, after some algebra, we obtain the optimal path:
\begin{numcases}
{{x(t)} =}\frac{3 A (1-2 |t|) \left(f+2 \tau ^3-4 \tau ^2+f \tau ^2+2 f \tau ^2 |t| -6 f |t|-8 \tau ^2 |t|
   +12  \tau |t| \right)}{(1-\tau )^3 \tau  (\tau +3)}, & $\tau/2<|t|<1/2$, \label{xoutside} \\
\frac{3 A \left(f \tau -\tau ^3-4 f \tau t^2  -8 f t^2+12 \tau t^2
   \right)}{(1-\tau ) \tau ^2 (\tau +3)},& $|t|<\tau/2$ , \label{xinside}
\end{numcases}
where we again introduced the area fraction $f=a/A$, $0<f<1$.  The left panel of Fig.~\ref{combined} shows, for $\tau=1/2$, the optimal paths for three values of the area fraction: $f=2/5$, $11/16$, and $4/5$.

\begin{figure}[h]
\includegraphics[width=0.32\textwidth,clip=]{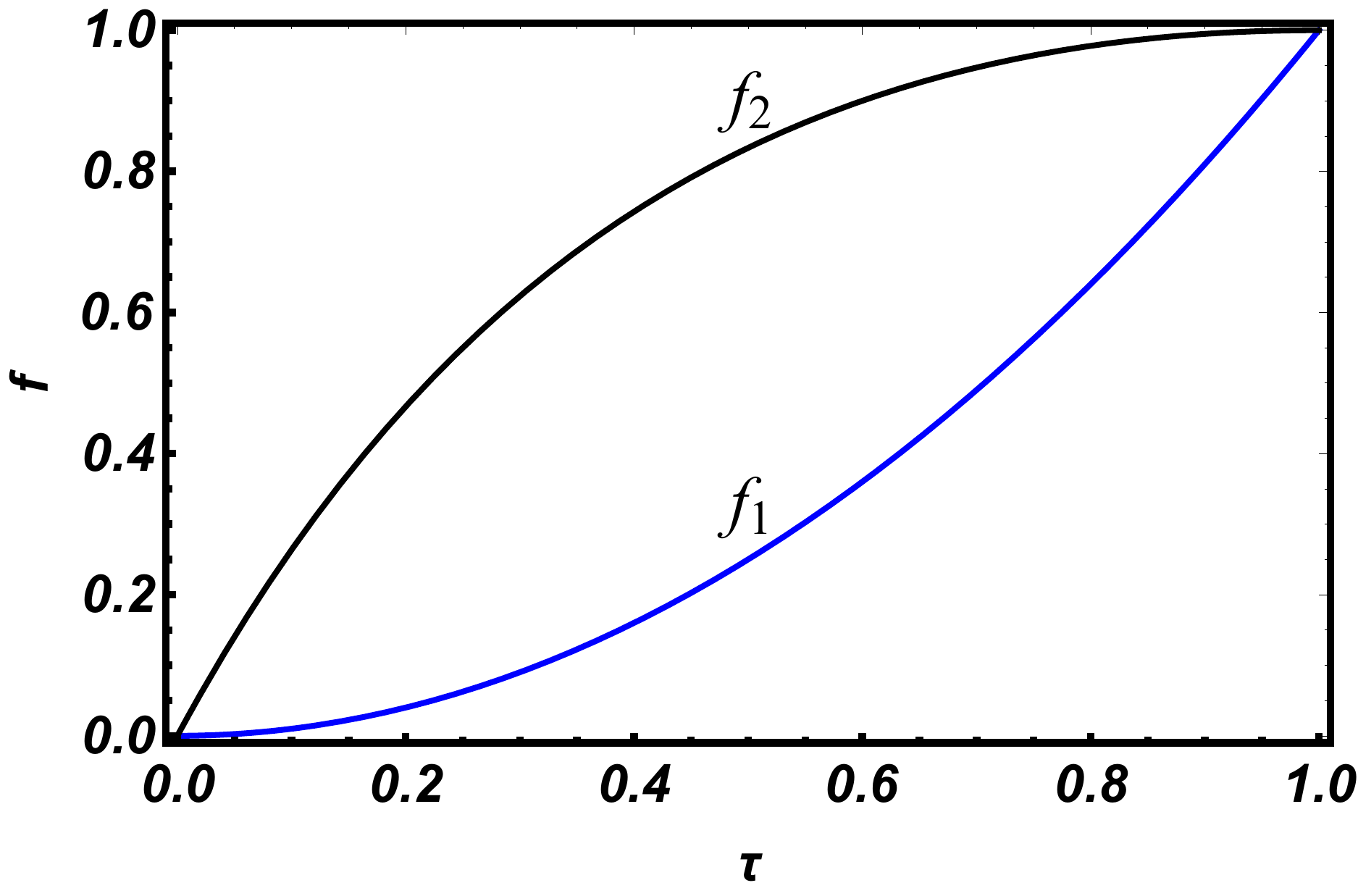}
\caption{The phase diagram for the large-$A$ and large-$a$ tail of $P(a,A,\tau)$. For $f_1<f<f_2$ the optimal paths $x(t)$ are given by Eqs.~(\ref{xoutside}) and~(\ref{xinside}) and illustrated in the left panel of Fig.~\ref{combined}. For $f<f_1$ they are given by Eqs.~(\ref{xoutsideonevoid})-(\ref{delta}) and illustrated in the middle panel of Fig. \ref{combined}. For $f>f_2$ they are given by Eqs.~(\ref{zerotwovoids})-(\ref{Delta}) and illustrated in the right panel of Fig.~\ref{combined}. The lines $f=f_1(\tau)$ and $f=f_2(\tau)$ are the lines of  second-order and third-order dynamical phase transitions, respectively.}
\label{phasediagram}		
\end{figure}

However, Eqs.~(\ref{xoutside}) and~(\ref{xinside}) describe legitimate Brownian excursions only when the area fraction $f$ satisfies the double inequality $f_1(\tau)\leq f\leq f_2(\tau)$, where
\begin{equation}\label{f1f2}
f_1 = \tau^2 ,\quad f_2 = \frac{\tau (3-\tau )}{1+\tau} .
\end{equation}
That is, Eqs.~(\ref{xoutside}) and~(\ref{xinside}) hold only in the region of $f$ between the two curves $f_1(\tau)$ and $f_2(\tau)$ on the phase diagram shown in Fig.~\ref{phasediagram}. At $f<f_1(\tau)$, $x(t)$ from Eq.~(\ref{xinside}) would become negative, which is forbidden, on a time interval around $t=0$. In its turn, at $f>f_2(\tau)$ $x(t)$ from Eq.~(\ref{xoutside})  would become negative on two time intervals adjacent to
$t=-1/2$ and $t=1/2$, respectively. The correct solutions at $f<f_1(\tau)$ and $f>f_2(\tau)$ are provided by the tangent construction of
the calculus of one-sided variations \cite{onesided} that we now present.

Let us start with $f<f_1(\tau)$. Here the correct optimal path $x(t)$ includes an interval $|t|<\delta/2$ around $t=0$, where $x(t)\equiv 0$. The length of this interval $\delta\leq\tau$ can be found from the continuity of $x(t)$ and its derivative $\dot{x}(t)$ at $t = \pm \delta/2$, and we finally obtain
\begin{numcases}
{{x(t)} =}\frac{3 A \left(1-\sqrt{f}\right) (1-2 |t|) \left(\sqrt{f}-2 \tau
-2\sqrt{f} |t|+4 |t| \right)}{(1-\tau
   )^3}, & $\tau/2<|t|<1/2$ , \label{xoutsideonevoid} \\
     \frac{3 A \left(1-\sqrt{f}\right) \left(\sqrt{f}-\tau -2 \sqrt{f}
   |t|+2 |t| \right)^2}{\sqrt{f} (1-\tau
   )^3},& $\delta/2<|t|<\tau/2$ , \label{xinsideonevoid}\\
     0,& $|t|<\delta/2$ ,    \label{zeroonevoid}
\end{numcases}
where
\begin{equation}\label{delta}
\delta=\frac{\tau-\sqrt{f}}{1 - \sqrt{f}} .
\end{equation}
When $f \to 0$, $\delta$ approaches $\tau$, and $x(t)$ vanishes on the whole subinterval $|x|<\tau$. The middle panel of Fig.~\ref{combined} shows an example, for $\tau=1/2$, of the optimal path $x(t)$ as described by Eqs.~(\ref{xoutsideonevoid})-(\ref{delta}) for the area fraction $f=1/8$, which is smaller than $f_1(\tau=1/2)=1/4$.

At $f>f_2(\tau)$ the optimal path $x(t)$ includes two symmetric intervals, where $x(t)\equiv 0$:  $\Delta/2<|t|<1/2$, where $\tau<\Delta\leq1$.   The parameter $\Delta$ can be found from the continuity of $x(t)$ and its derivative $\dot{x}(t)$ at $t = \pm \Delta/2$. As a result, $x(t)$ in this regime is the following:
\begin{numcases}
{{x(t)} =} 0, & $\Delta/2<|t|<1/2$ , \label{zerotwovoids} \\
   \frac{6 A (1-f) f \left(4 f |t|+f \tau -\sqrt{(9-f) (1-f)} \tau -3 \tau
   \right)^2}{\left(3-3 f+\sqrt{(9-f) (1-f)}\right)^3 \tau ^3}, & $\tau/2<|t|<\Delta/2$ , \label{xoutsidetwovoids}\\
    \frac{6 A (1-f) f \left(3 \tau^2 -f \tau ^2
    +\sqrt{(9-f) (1-f)} \tau ^2 -8 f t^2\right)}{\left(3-3 f+\sqrt{(9-f) (1-f)}\right)^2 \tau ^3},& $|t|<\tau/2$ ,    \label{xinsidetwovoids}
\end{numcases}
where
\begin{equation}\label{Delta}
\Delta = \frac{\left(3-f+\sqrt{(9-f) (1-f)}\right) \tau }{2 f} .
\end{equation}
When $f \to 1$, $\delta$ approaches $\tau$, and $x(t)$ vanishes on the external subintervals $\tau/2<|t|<1/2$. The right panel of Fig.~\ref{combined} depicts an example, for $\tau=1/2$, of the optimal path $x(t)$ described by Eqs.~(\ref{zerotwovoids})-(\ref{Delta}) for the area fraction $f=0.95$, which is larger than $f_2(\tau=1/2)=5/6$.

Having determined the optimal path $x(t)$ for all $0<f<1$, we can now compute the action $s(f,A,\tau)$, which describes the large-$a$ and $A$ tail of $P(a,A,\tau)$ up to a pre-exponential factor: $-\ln P(a,A,\tau) \simeq s(f,A,\tau)$.  We use Eq.~(\ref{sA}) (with the already known $\lambda_1$ and $\lambda_2$) separately in the regimes $0<f<f_1$, $f_1<f<f_2$ and $f_2<f<1$. In each regime we subtract from the result the action $s_0(A)=6 A^2$ of the right tail of the Airy distribution in order to account for the denominator in Eq.~(\ref{P(aA)}). After some algebra, we arrive at the following expressions, which cover the whole range  $0<f<1$:
\begin{numcases}
{-\ln P(a\equiv f A,A,\tau)\simeq s(f,A,\tau)=} \left[\frac{24 \left(1-\sqrt{f}\right)^2}{(1-\tau )^3}-6\right] A^2 , & $0<f<f_1$ , \label{sfsmall} \\
  \frac{6 \left(2 f+\tau ^3-3 \tau \right)^2}{(1-\tau
   )^3 \tau ^2 (\tau +3)} \,A^2, & $f_1<f<f_2$ , \label{sfintermediate}\\
  \left[\frac{3 \left(27-f^2-(9-f)^{3/2}(1-f)^{1/2} -18 f\right)}{4 \tau ^3}-6\right] A^2,& $f_2<f<1$ .    \label{sflarge}
\end{numcases}

\begin{figure}[h]
\includegraphics[width=0.40\textwidth,clip=]{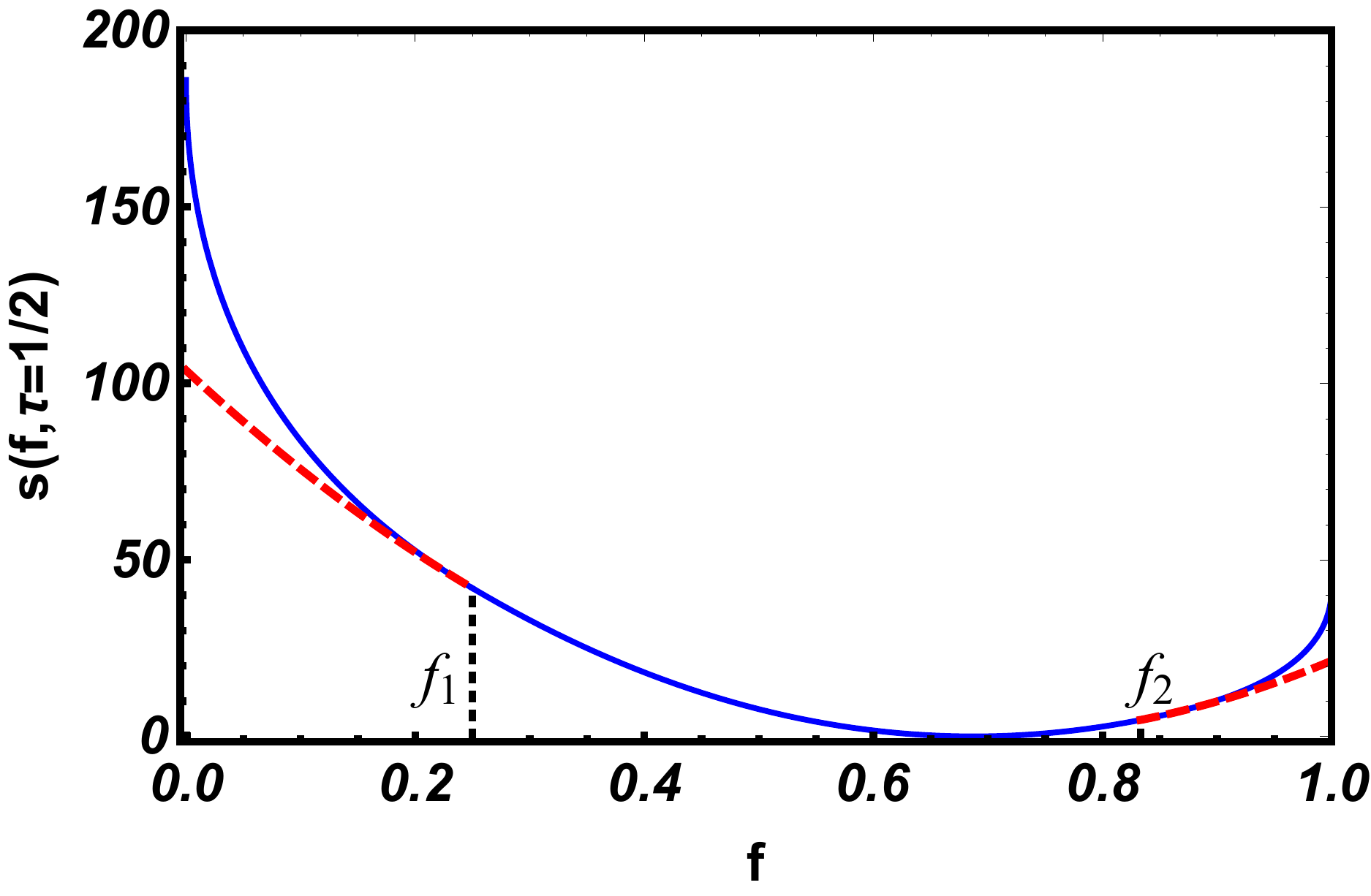}
\caption{The large-deviation function $s(f,\tau,A)$, Eqs.~(\ref{sfsmall})-(\ref{sflarge}), which describes the large $a$ and large $A$ tail of $P(a,A,\tau)$:  $-\ln P(a,A,\tau) \simeq s(f=a/A,\tau,A)$.  Shown is $s(f,\tau,A)/A^2$ versus $f$ for $\tau=1/2$. There are second- and third-order dynamical phase transitions at the points $f=f_1$ and $f=f_2$, respectively. The red dashed lines shows an illegitimate extension of Eq.~(\ref{sfintermediate}) onto the whole interval $0<f<1$. Although this extension has a smaller action outside of the interval $(f_1,f_2)$, it is incorrect because the corresponding path $x(t)$ crosses the origin thus violating the excursion condition. Notice that Eq.~(\ref{sfintermediate}) would be correct on the whole interval $0<f<1$ if we considered a Brownian bridge, rather than a Brownian excursion.}
\label{s(ftau)}		
\end{figure}

One conspicuous feature of the  large deviation function $s(f,A,\tau)$ in this limit is its quadratic scaling $A^2$, as in the Airy distribution at large $A$. At fixed $A$ and $\tau$ the function $s(f,A,\tau)$ has a minimum, equal to zero, at $f=f_{\text{opt}}=(1/2) \tau  \left(3-\tau ^2\right)$ which  lies in the region $f_1<f<f_2$, see Fig.~\ref{s(ftau)}.  At this special value of $f$ the optimal path $x(t)$ is described, for all $|t|<1/2$, by a single parabola $x(t)=(3A/2) (1-4t^2)$, which corresponds to the large-$A$ tail of the Airy distribution, unconstrained by the additional condition (\ref{a}).  In the region of $f_1<f<f_2$ the function $s$ is a quadratic function of $f$, which corresponds to Gaussian fluctuations of $a$ at large $A$. Outside of the region $f_1<f<f_2$ the fluctuations of $a$ are non-Gaussian.

Importantly, the function $s(f,A,\tau)$  vs. $f$ is non-analytic along the curves $f=f_1(\tau)$ and $f=f_2(\tau)$, where the character of the optimal path changes, see Fig.~\ref{phasediagram}.  At fixed $\tau$ and $f=f_1$, $s$ is continuous together with its first derivative with respect to $f$. The second derivative with respect to $f$, however, experiences a jump which can be interpreted as a second-order dynamical phase transition. At fixed $\tau$ and $f=f_2$, $s$ is continuous together with its first and second derivatives with respect to $f$. Here the third derivative has a jump, so this is a third-order transition.
Similar in spirit dynamical phase transitions, predicted by  geometrical optics in conjunction with the calculus of one-sided variations,  have been recently reported in a series of works on Brownian motions, pushed into large deviation regimes by constraints \cite{Meerson2019,SmithMeerson2019a,SmithMeerson2019b,3short,MMajumdar2020}.

One more interesting feature of $s(f,A,\tau)$ is observed when we set $f=(X/A) \tau$ and pass to the limit of $\tau \to 0$ while keeping $X/A=\text{const}$. This limits probes (the large-$A$ tail of) the distribution of the excursion position $X$ at time $t=0$, conditioned on the total area $A$. The latter distribution has been recently studied in Ref. \cite{Agranov2020}.  From the expressions~(\ref{f1f2}) for $f_1(\tau)$ and $f_2(\tau)$ we see that, as $\tau$ goes to zero, $f_1<f<f_2$ for $X/A<3$ and $f>f_2$ for $X/A>3$. Plugging $f=(X/A) \tau$ into Eqs.~(\ref{sfintermediate}) and~(\ref{sflarge}) and taking the limit of $\tau \to 0$, we arrive at
\begin{numcases}
{s(X,A)=} 8 \left(X/A-3/2\right)^2, & $X/A<3$ , \label{below} \\
(8/9)(X/A)^3-6, & $X/A>3$ , \label{above}
\end{numcases}
in perfect agreement with Ref.~\cite{Agranov2020}.  As it was noticed in Ref.~\cite{Agranov2020}, the large deviation function $s(X,A)$ exhibits a third-order transition at $X/A=3$. Now we see that the third-order transition in $P(a,A,\tau)$, that we discussed above, is preserved in the limiting procedure which restores the position distribution from the (more general) area distribution on the subinterval at given (and very large) total excursion area.

Now we can recap our main results for Model 2.   The behaviors of the distribution $P(a,A,\tau)$ at small $a$ and $A$ and large $a$ and $A$   are described by Eq.~(\ref{ratefunctionr}) and Eqs.~(\ref{sfsmall})-(\ref{sflarge}), respectively.   The large-deviation function $s(f,A,\tau)$ in Eqs.~(\ref{sfsmall})-(\ref{sflarge}) exhibits two dynamical phase transitions -- of the second and third order -- at $f=f_1$ and $f=f_2$, respectively.

\section{Summary and Discussion}
\label{summary}

In Model 1 we studied fluctuations of the area $a$ under the curve, describing the position of a Brownian excursion as a function of time, on a subinterval of the excursion. For simplicity, we centered the subinterval at $t=T/2$.  We obtained an exact integral expression (\ref{LaplaceP}) for the Laplace transform of the probability distribution of $a$, and extracted the small-$a$ and large-$a$ tails of the distribution. For the small-$a$ tail we also succeeded in evaluating the pre-exponential factors which reveal the existence of two different asymptotic regimes when the length of the subinterval $\tau$ is very close to $T$, the total duration of the excursion.

\begin{figure}[h]
\includegraphics[width=0.40\textwidth,clip=]{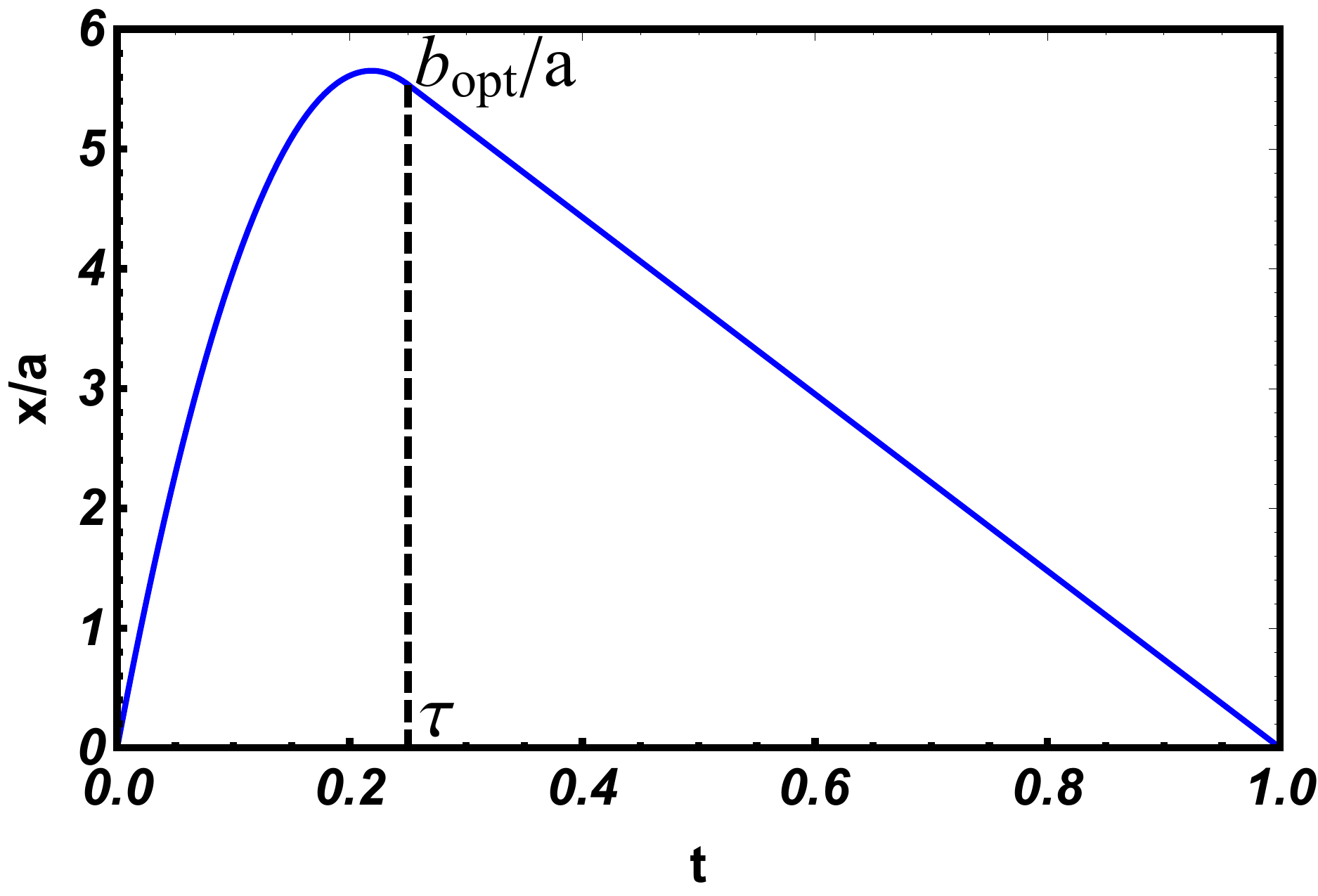}
\caption{Rescaled optimal path $x(t)/a$ of the Brownian excursion on the interval $0<t<1$, conditioned on the area $a$ on the subinterval $0<t<\tau$. The parabolic segment of $x(t)$ on $0<t<\tau$ continuously matches, together with its derivative, with the ballistic segment on $\tau<t<1$. In this example $\tau=1/4$.}
	\label{rambeaux}		
\end{figure}

Model 1 is very similar to the model proposed and studied earlier (by the same path-integral method) by Rambeau and Schehr \cite{Rambeau2009}. Their model also deals with a Brownian excursion on $0<t<T$ and studies the distribution of the area on  a sub-interval of duration $\tau<T$, but in their case the subinterval is $t\in (0, \tau)$. Their distribution has the same scaling behavior~(\ref{pa1}), but the scaling function is different. The distribution tails were not addressed in Ref.~ \cite{Rambeau2009}. The small-area tail can be extracted from Eq.~(69) of Ref. \cite{Rambeau2009} along the lines of our Sec.~\ref{lefta}, that is by keeping only the $k=1$ term of the series in $k$   and performing a saddle-point evaluation of the inverse Laplace transform. Up to a pre-exponent, this tail has a Donsker-Varadhan large-deviation form, and it coincides with the corresponding tail (\ref{DVsmalla}) of Model 1. This coincidence is not surprising since, in this leading-order approximation, only the subinterval duration contributes, while the location of the subinterval inside the interval $0<t<T$ is irrelevant.

The large-area tail of the Rambeau-Schehr distribution  can be described by geometrical optics along the lines of our Sec.~\ref{righta}.  This leading-order calculation yields a Gaussian tail
\begin{equation}\label{righttailRS}
- \ln P(a,\tau) \simeq
\frac{6 a^2}{\tau ^3 (4-3 \tau )}\,,
\end{equation}
which differs from the right tail of Model 1, see Eq.~(\ref{righttailonlya}). That is, the right tail strongly depends on the exact location of the subinterval.  In particular, as $\tau \to 0$,  the essential singularity $P(a,\tau) \sim \exp(-2a^2/\tau^2)$, observed in Model 1 [see Eq. (\ref{righttailonlya})], is replaced by a stronger essential singularity
$P(a,\tau) \sim \exp(-3a^2/2\tau^3)$ in the model of Rambeau and Schehr \cite{Rambeau2009}. The difference can be intuitively understood: it is much less probable to observe a large subinterval area if the subinterval is adjacent to the starting point $t=0$ where $x=0$. An example of the optimal path, corresponding to Eq.~(\ref{righttailRS}), is shown in Fig. \ref{rambeaux}. When $\tau=1$,  both Eq.~(\ref{righttailonlya}) and Eq.~(\ref{righttailRS}) yield the right tail of the Airy distribution, $- \ln P(a,\tau=1)\simeq 6a^2$, see Eq.~(\ref{high}).

In Model 2 we studied fluctuations of the area $a$ on the subinterval, conditioned on the total area $A$ under the excursion. Here we found the leading-order small-area and large-area asymptotics of the distribution. Model 2 turns out to be richer in its behavior than Model 1. Here the large-area asymptotic exhibits two dynamical phase transitions, which result from changes in the character of the optimal path of the conditioned Brownian excursion at critical values of the area fraction $f=a/A$. These phase transitions are direct consequences of the excursion restriction $x(t)>0$ for all $|t|<T$, and they can be described by geometrical optics of Brownian motion in conjunction with the calculus of one-sided variations.

Finally, it would be interesting to test our results for Model 1 and Model 2 experimentally in a dilute colloidal system, extending the method of Ref. \cite{Agranov2020}, where the Airy distribution was measured.

\section*{Acknowledgments} I thank Tal Agranov for useful comments and for help with Fig. 2. I am very grateful to Naftali R. Smith for a critical reading of the manuscript.
This work was supported by the Israel Science Foundation (Grant No. 807/16).

\end{document}